\documentclass[twocolumn,prb,aps,superscriptaddress,showpacs]{revtex4}

\usepackage{dcolumn}
\usepackage{amsmath}
\usepackage{epsfig}

\newlength{\figwidth}
\newlength{\figwidthb}
\begin{document}

\title{Electronic Structure of Doped Lanthanum Cuprates Studied with Resonant Inelastic X-Ray Scattering}
\author{D. S. Ellis}
\affiliation{Department of Physics, University of Toronto, Toronto,
Ontario M5S~1A7, Canada}
\affiliation{Materials Dynamic Laboratory, SPring-8, RIKEN
1-1-1 Kouto, Sayo, Hyogo, 679 Japan}
\author{Jungho Kim}
\affiliation{Department of Physics, University of Toronto, Toronto,
Ontario M5S~1A7, Canada}
\affiliation{XOR, Advanced Photon Source, Argonne National
Laboratory, Argonne, Illinois 60439}
\author{Harry Zhang}
\affiliation{Department of Physics, University of Toronto, Toronto,
Ontario M5S~1A7, Canada}
\author{J. P. Hill}
\affiliation{Condensed Matter Physics \& Materials Science Department, Brookhaven National Laboratory,
Upton, New York 11973}
\author{Genda Gu}
\affiliation{Condensed Matter Physics \& Materials Science Department, Brookhaven National Laboratory, Upton, New York 11973}
\author{Seiki Komiya}
\affiliation{Central Research Institute of Electric Power Industry, Yokosuka, Kanagawa 240-0196, Japan}
\author{Yoichi Ando}
\affiliation{Institute of Scientific and Industrial Research, Osaka University, Ibaraki, Osaka 567-0047, Japan}
\author{D. Casa}
\affiliation{XOR, Advanced Photon Source, Argonne National
Laboratory, Argonne, Illinois 60439}
\author{T. Gog}
\affiliation{XOR, Advanced Photon Source, Argonne National
Laboratory, Argonne, Illinois 60439}
\author{Young-June Kim}
\email{yjkim@physics.utoronto.ca}
\affiliation{Department of
Physics, University of Toronto, Toronto, Ontario M5S~1A7, Canada}

\date{\today}

\begin{abstract}

We report a comprehensive Cu $K$-edge RIXS investigation of  $\rm La_{2-x}Sr_xCuO_4$ (LSCO) for 0$\leq$x$\leq$0.35, stripe-ordered $\rm La_{1.875}Ba_{0.125}CuO_4$ (LBCO), and $\rm La_{2}Cu_{0.96}Ni_{0.04}O_4$ (LCNO) crystals. The RIXS spectra measured at three
high-symmetry momentum transfer (\textbf{q}) positions are compared as a function
of doping and for the different dopants. The spectra in the energy range 1-6 eV can be described with three broad peaks, which evolve systematically with increased doping.  The most systematic trend was observed for \textbf{q}=($\pi$, 0) corresponding to the zone boundary. As hole doping increased, the spectral weight transfer from high energies to low energies is nearly linear with \emph{x} at this \textbf{q}.  We interpret the peaks as interband transitions in the context of existing band models for this
system, assigning them to Zhang-Rice band$\rightarrow$upper Hubbard band, lower-lying band$\rightarrow$upper Hubbard band, and lower-lying band$\rightarrow$Zhang-Rice band transitions.    The spectrum of stripe-ordered LBCO was also measured, and found to be identical to the correspondingly doped LSCO, except for a relative enhancement of the near-infrared
peak intensity around 1.5-1.7 eV. The temperature dependence of this near-infrared peak in LBCO was more pronounced than for other parts of the spectrum, continuously decreasing in intensity as the temperature was raised from 25 K to 300 K.  Finally, we find that 4\% Ni substitution in the Cu site has a similar effect on the spectra as does Sr substitution in the La site.
\end{abstract}


\maketitle

\section{Introduction}

The strongly interacting nature of valence electrons in systems such as the lamellar copper oxides gives rise to exotic phases such as high-temperature superconductivity\cite{Bednorz88} and charge stripes.\cite{Tranquada95} This has spurred much study of the electronic structure of these materials.  The electronic structure is often described as consisting of an upper Hubbard band above the Fermi level, oxygen states and lower Hubbard band states below, and the hybridized Zhang-Rice singlet states \cite{Zhang88} near the Fermi level.  A variety of spectroscopic techniques have been used to probe the electronic states.  Below the Fermi level, the electronic band structure of the ground state has been well characterized by angle resolved photoemission spectroscopy (ARPES).\cite{Ino02,Yoshida03,Yoshida06}  In those studies, lightly doped $\rm La_{2-x}Sr_xCuO_4$ (LSCO)  was shown to have a Fermi arc in the vicinity of ($\pi$/2 $\pi$/2) in momentum space.  Furthermore, a large hole-like Fermi surface centered at ($\pi$, 0) appearing as a wide flat band approaches the Fermi-level from below at $x\sim$0.12.  Recently, quantum oscillations\cite{DorionLeyraud07} and Hall coefficients\cite{LeBoeuf07} studies in related compounds have suggested the existence of electron-like patches of Fermi surface not previously detected before with ARPES.\cite{quantum_osc_note}\\

The electronic states above the Fermi level have been accessed through two-particle spectroscopies such as optical Raman scattering\cite{Lyons88,Sugai88,Salamon95}, optical conductivity,\cite{Uchida91,Uchida96} third harmonic generation,\cite{Kishida03} and electron energy loss spectroscopy.\cite{Fink94}  Perhaps the most comprehensive study of the evolution of the electronic excitation spectrum with doping in LSCO  is Uchida et al.'s study of optical conductivity.\cite{Uchida91}  The parent compound, $\rm La_2CuO_4$, is a charge-transfer insulator,\cite{Zaanen85} with a charge transfer (CT) gap of $\sim$2 eV.  As holes are doped into  $\rm La_2CuO_4$ by strontium substitution into the lanthanum sites,  the spectral weight of the optical conductivity above the insulating gap sharply decreases.  At the same time, a low energy Drude intensity as well as ``in-gap" features including a mid-infrared (MIR) peak around 0.5 eV, and a near-infrared (NIR) peak at 1.5 eV emerge.  These observations seem to be common to LSCO, bilayer $\rm YBa_{2}Cu_{3}O_{6+y}$, and electron-doped $\rm Nd_{2}CuO_{4-y}$.\cite{Thomas92,Basov05} The origin of the MIR peak is still for the most part uncertain, although it is thought to be
intimately linked to high-temperature superconductivity\cite{Leggett99,Lee05}, and a MIR peak in $\rm La_2CuO_4$ has been associated with multi-magnon excitation\cite{Lyons88}.  The NIR spectrum, on the other hand, had not drawn much attention until recently.  When first observed, the 1.5 eV peak was thought to be of extrinsic origin \cite{Uchida91}.  Measurements of this peak in samples that were hole-doped by oxygen,\cite{Quijada95} photodoping measurements,\cite{Ginder88} and a comparison between materials \cite{Waku04} have together demonstrated that the 1.5 eV feature comes solely from the presence of holes in the copper-oxygen plane, although this last study extended only to $x$=0.10. \\

\vspace{-3 mm}

Resonant inelastic x-ray scattering (RIXS) is capable of measuring element-specific electronic excitations at finite momentum transfer and has been used to study the cuprates and other transition metal compounds.  Specifically,  RIXS at the Cu $K$-edge energy has been used extensively to observe excitations in $\rm La_2CuO_4$,\cite{Abbamonte99,Kim02,Collart06,Lu06,Ellis08} and the doping dependence has been studied to some degree.\cite{Kim04,Wakimoto05,Collart06}  In addition to improvements in instrumentation, recent years have seen advances in understanding of the relation of the RIXS cross-section to the optical loss function and the dynamical structure factor, through both empirical\cite{Kim09} and theoretical means\cite{Ament07}.  Theorists have also been engaged in understanding doping-dependent RIXS spectra in terms of electronic band structure,\cite{Tsutsui99, Markiewicz06b} although a detailed comparison between theory and experiment has been limited to only a few dopings.\\

In fact, a sizeable body of RIXS data of $\rm La_{2-x}Sr_xCuO_4$ has been accumulating in the literature.  A comparison of the RIXS spectra for \emph{x}=0, \emph{x}=0.05 and \emph{x}=0.17 by Kim et.~al,\cite{Kim04} showed that while the spectra of the undoped and \emph{x}=0.05 samples were, for the most part, similar, the high-energy intensity for \emph{x}=0.17 decreased and shifted to still higher energies.  Wakimoto et.~al.\cite{Wakimoto05} compared the Cu $K$-edge RIXS spectra in overdoped samples to undoped samples, and found that the spectral weight above the charge transfer gap shifts to successively higher energies in the case of the overdoped samples.  For the low energy part of the spectra, a peak at around 1.5 eV for \emph{x}=0.30 was observed and found to have its highest intensity when the incident energy was between 8992 eV and 8993 eV.  Collart et.~al. \cite{Collart06} compared \emph{x}=0 and \emph{x}=0.07 samples, and noted that the sharp excitonic peaks above the charge transfer gap were greatly reduced in the doped sample, although the coarse instrumental resolution in that experiment prevented observation of the  NIR features.  In parallel studies using Cu $L$-edge RIXS, Ghiringhelli et al.,\cite{Ghiringhelli04}  observed a main peak between 1.5 eV and 2.0 eV in both undoped and optimally doped LSCO samples, and determined the energy to be consistent with their $dd$ excitation calculations.  Finally, in a recent RIXS study on stripe-ordered cuprate and nickelate crystals, an increase of the $\sim$1 eV intensity near the stripe ordering wavevector was observed and interpreted to be an anomalous softening of the charge excitations due to stripes.\cite{Wakimoto09}\\

In this paper we present comprehensive studies of the doping dependence of the Cu $K$-edge RIXS spectra for samples spanning a wide range of dopings.  Combined with previous data,\cite{Wakimoto05,Ellis08} systematic trends in the spectral weight shift from lower to higher energy, and changes in the individual peaks are observed.  We discuss these results in terms of inter-band transitions using simple models, and find that the intensity and peak position trends with doping are consistent with the expectation of a rigid band phenomenological model with a low-lying band, Zhang-Rice band and a upper Hubbard band.  The paper is organized as follows: Section \ref{sect:Experimental} outlines the experimental method.  The results are presented in Sec.~\ref{sect:Observations}, beginning with the wide-range spectra as a function of doping at different momentum transfers in Sec.~\ref{subsect: Overall doping dependence}.  Sec.~\ref{subsect:Near-IR Feature} focuses on the near-infrared region, including fitting results for the near-infrared peak.  The spectra of the $x$=1/8 Ba-doped sample, and its temperature dependence, are presented in Sec.~\ref{subsect:Ba doped}.  Analysis and discussion of these spectra, including spectral weights and individual peaks, comprise Sec.~\ref{sect: Discussion}.  Sec.~\ref{sect: Conclusion} summarizes our findings.\\

\vspace{-3 mm}

\section{Experimental Method}
\label{sect:Experimental}
Single crystals of LSCO for \emph{x}=0.05 and
0.17 were the same as used in Ref.~\onlinecite{Kim04}.  Additional crystals with dopings \emph{x}=0.07, 0.10 and 0.12 were grown by
the floating-zone method and were annealed at $900\,^{\circ}\mathrm{C}$ for 30 hours in
flowing oxygen. These samples were then characterized by measuring magnetization with a Quantum
Design SQUID magnetometer and had $\rm T_c$ of 14~K, 27.5~K and 30~K respectively.  A non-superconducting $x$=0.35 crystal and $\rm La_{2}Cu_{0.96}Ni_{0.04}O_4$ (LCNO) with N\'eel temperature $\rm T_N$=317 K was similarly grown.  In addition, we measured a $x$=$\frac{1}{8}$ stripe-ordered $\rm La_{2-x}Ba_xCuO_4$ (LBCO) crystal, the same sample used in Ref.~\onlinecite{Kim08}.  The $x$=0.25 and $x$=0.30 LSCO samples were studied in Ref.~\onlinecite{Wakimoto05} and the data were measured with $\sim$240 meV energy resolution.  A summary of the samples is presented in Table~1. \\

\begin{table}

\caption {Table of samples whose spectra are included in this study.  Listed from left to right : Dopant concentration $x$ or $y$ in $\rm La_{2-x}(Sr/Ba)_xCu_{1-y}Ni_yO_4$, superconducting $T_c$, surface normal direction (tetragonal notation), where the sample was grown (University of Toronto, Massachusetts Institute of Technology, Brookhaven National Laboratory, or Central Research Institute of Electric Power Industry).}

\begin{ruledtabular}

\begin{tabular}{ccccc}

Doping (x) & $T_c$ (K) & Orientation & Grown & Note \\

\hline

0.00 & N/A & (100) & MIT & Ref.\onlinecite{Ellis08}\\

0.05 & N/A & (110) & CRIEPI & this work\\

0.07 & 14 & (100) & Toronto & this work\\

0.10 & 27 & (100) & Toronto & this work\\

0.12 & 30 & (100) & Toronto & this work\\

0.17 & 42 & (100) & CRIEPI & this work\\

0.25 & 15 & (100) & Toronto & Ref.\onlinecite{Wakimoto05}\\

0.30 & $<$ 2 & (100) & Toronto & Ref.\onlinecite{Wakimoto05}\\

0.35 & N/A & (100) & Toronto & this work\\

\hline
4\% Ni & N/A & (100) & Toronto & this work\\

\hline
1/8 Ba & 6 & (100) & BNL & this work\\

\end{tabular}

\end{ruledtabular}

\end{table}

The RIXS measurements were all done at the 9ID beamline at the Advanced Photon Source.  The incident polarization in all cases was parallel to the crystallographic $c$-axis, and the peak energy of Cu $K$-edge fluorescence was chosen as the incident energy. \cite{Kim02}  It is known that the resonant enhancement occurs when the incident energy is near the absorption peak for both doped and undoped samples, and is more than 1 eV broad in incident energy.\cite{Kim02,Kim04,Wakimoto05,Lu06}    In our notation, the two-dimensional reduced momentum transfer \textbf{q}$\equiv$ \textbf{Q}-\textbf{G}, where \textbf{Q} is the change in wave-vector of the x-ray upon reflection from the sample, and \textbf{G} is the closest reciprocal lattice vector.  RIXS data were obtained for \textbf{q} about the Brillouin zone-center (3 0 0), focusing on \textbf{q}=($\pi$,0) (zone boundary along the Cu-O bond direction), \textbf{q}=($\pi$,$\pi$) (zone boundary at a 45$^{\circ}$ angle), and \textbf{q}$\leq$0.2$\cdot$($\pi$,0).  To avoid intense quasi-elastic background, \textbf{q} within the first fifth of the Brillouin zone -- but not exactly at (3 0 0) -- was used as an approximation for the zone center,\cite{Ellis08} since the momentum resolution of our experiments was about 10-20\% of the Brillouin zone.  The reference point for zero energy loss, $\hbar\omega\equiv$ E$_i$-E$_f$ was set to the elastic peak of each spectra.  The incident energy, E$_i$, was kept fixed at the copper $K$-edge absorption peak ($\sim$8993 eV) and the final energy E$_f$ was scanned.  Depending on whether a 1-meter or 2-meter detector arm was used and on the detector slit size, the overall energy resolution varied: $\sim$150 meV for measurements on the \emph{x}=0.05, \emph{x}=0.10, \emph{x}=0.17 and LCNO samples, and 200-300 meV for the \emph{x}=0.07 and \emph{x}=0.12 samples, and $\sim$350 meV for \emph{x}=0.35.\\

All measurements were performed at low temperature (25K or less) using a closed-cycle He refrigerator except for \emph{x}=0.12 and \emph{x}=0.35 which were performed at room temperature.  A subsequent experiment on the \emph{x}=0.12 sample was performed at low temperature, and for this measurement we used a pixel-array detector with a 1-m detector arm, to measure many energies simultaneously\cite{Huotari05} with a resolution of 130 meV.  This same experimental setup was also used for the study of the LBCO crystal.\\

%=======================================================================================
\begin{figure*}
\begin{center}
\epsfig{file=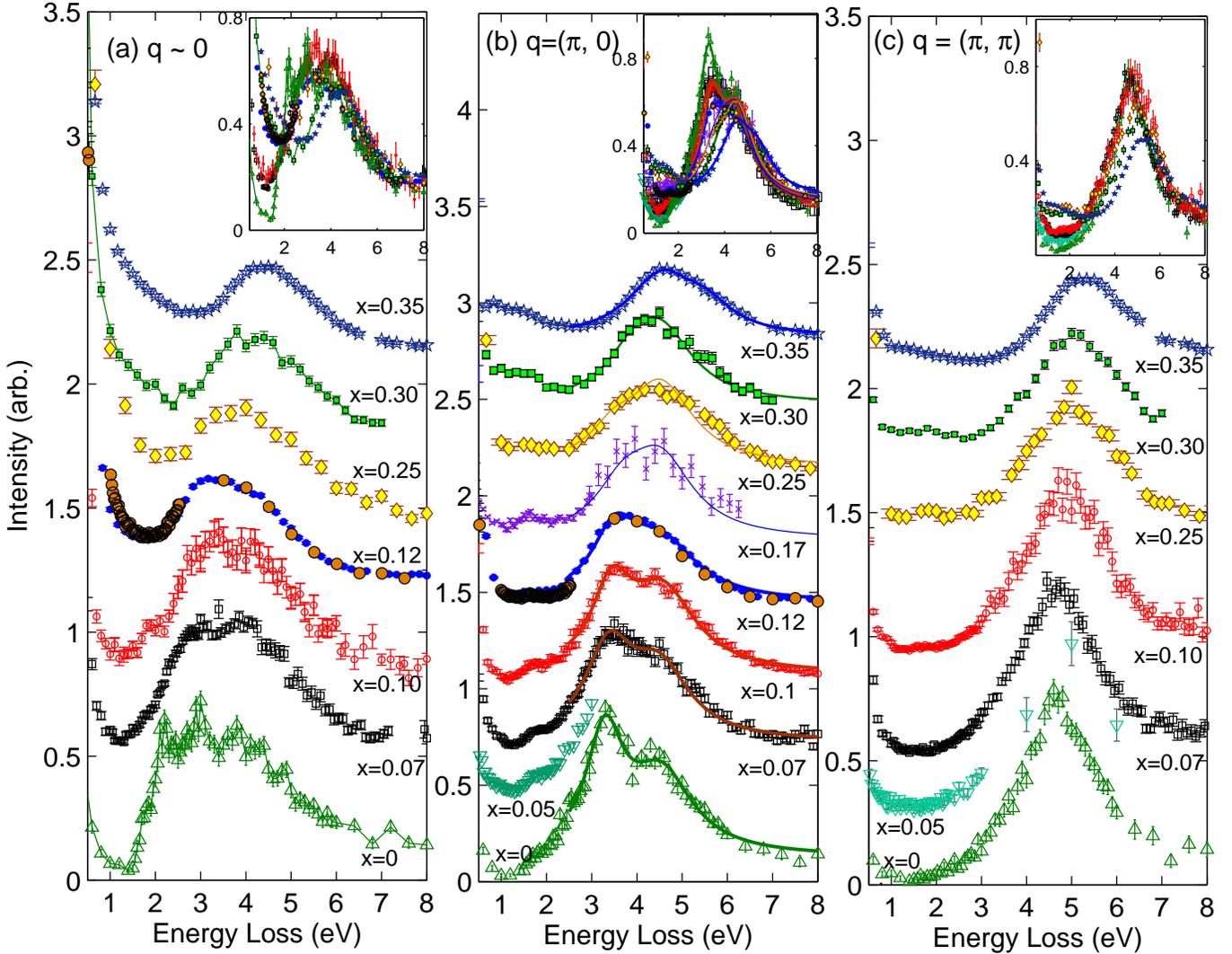,height=5.7in,keepaspectratio}
\end{center}
\caption{(Color Online) Progression of RIXS spectra of LSCO at momentum transfers corresponding to (a) near zone center (b) the zone boundary \textbf{q}=($\pi$,0), and (c) zone boundary \textbf{q}=($\pi$,$\pi$), as the sample doping is increased.  The curves are vertically shifted with respect to each other for clarity.  For x=0.l2, the bigger orange circles with black outlines, with less point density at higher energy loss, are from the higher resolution measurement, as described in the text.  The insets show the combined, unshifted curves overlayed with each other.  The data for \emph{x}=0.30 and \emph{x}=0.25
are from Ref.~\onlinecite{Wakimoto05} while the \emph{x}=0 data are
from Ref.~\onlinecite{Ellis08}.  The data were normalized as specified in the text.  The solids lines in (b) are fits to two Lorentzians.} \label{fig:widerange}
\end{figure*}
%=======================================================================================

\section{Experimental Results}
\label{sect:Observations}

\subsection{Overall doping dependence}
\label{subsect: Overall doping dependence}

Figure~\ref{fig:widerange} shows the doping dependent evolution of the RIXS spectra for LSCO at reduced momenta \textbf{q}$\approx$(0,0), \textbf{q}=($\pi$,0), and \textbf{q}=($\pi$,$\pi$) respectively.  The spectra were normalized such that their respective intensities at around 10 eV match with each other.  This high-energy intensity is thought to come mostly from Cu $\rm K \beta_5$ fluorescence, which is proportional to the sample volume probed.\cite{Kim04}  For \emph{x}=0.17, we only took data to 7 eV, so the data was normalized to match the other curves at its highest energy.  In the case of \emph{x}=0.05, data was not measured up to high energy, so the data was scaled by eye to match the trend of progression of the other curves.  For clarity of the individual spectra, they are displaced with respect to each other along the intensity axis.  To see the evolution of the relative intensities as a whole,  the undisplaced spectra are overlayed on the same graph in the insets.\\

%=======================================================================================
\begin{figure}
\begin{center}
\epsfig{file=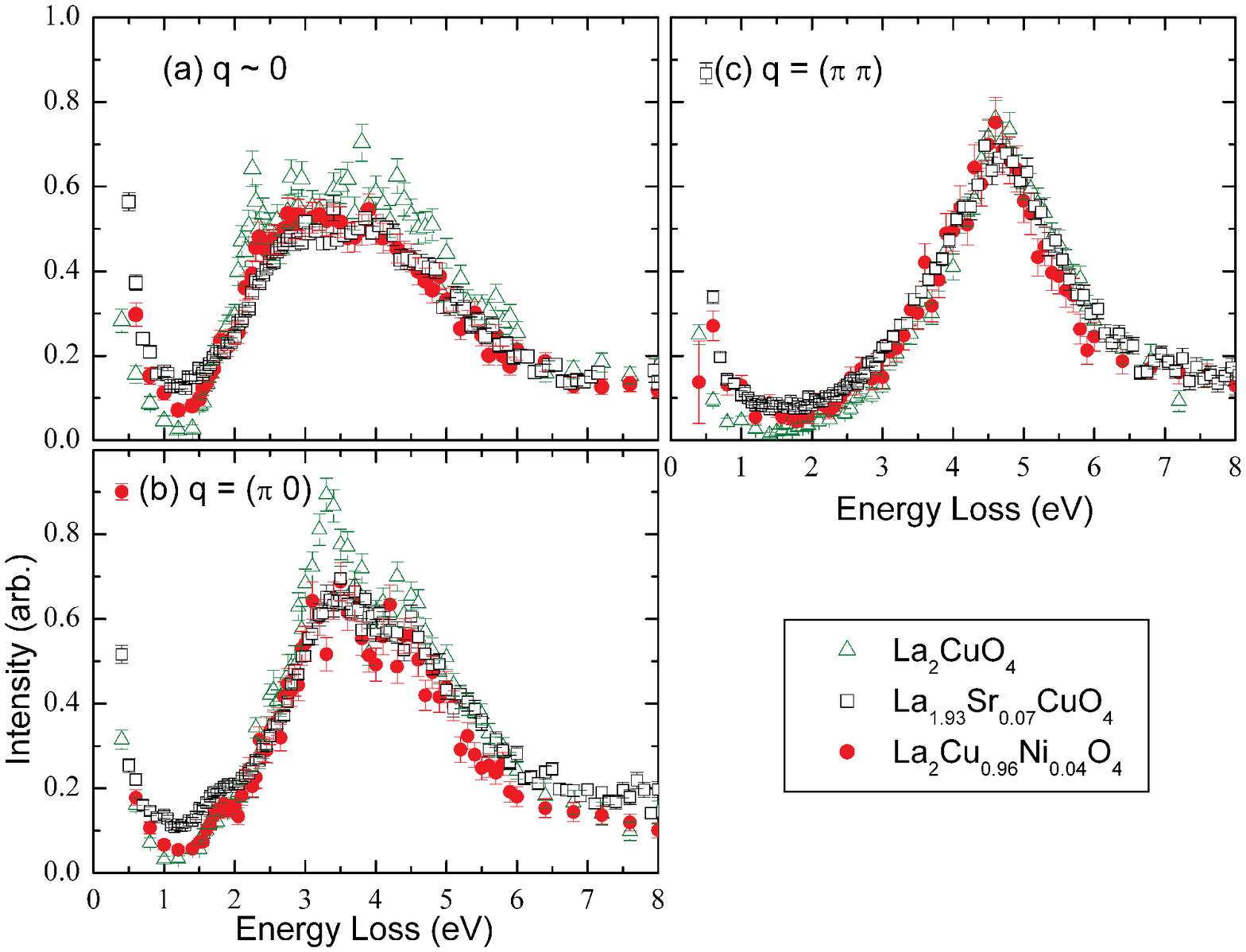,width=3.3in,keepaspectratio}
\end{center}
\caption{(Color Online) Comparison of the effect of light doping for LSCO and LCNO, at three momentum transfers : (a) near zone center, (b) \textbf{q}=($\pi$,0), and (c) \textbf{q}=($\pi$,$\pi$).} \label{fig:lcno}
\end{figure}
%=======================================================================================

For the \textbf{q}$\approx$(0 0) spectra of Fig.~\ref{fig:widerange}(a), the sharp peak at 2.2 eV present for $x$=0 decreases dramatically by $x$=0.07.  In fact, this sharp feature is only observed for the $x$=0.0 sample.  To investigate this further, we have studied Ni-doped $\rm La_2CuO_4$.  In Fig.~\ref{fig:lcno}, the RIXS spectra of 4\% Ni doped sample are compared with those of undoped and 7\% Sr doped samples.  Note that Ni-doping does not introduce charge carriers, but, rather the Ni dopant acts as magnetic impurity in the $\rm CuO_2$ plane.  Despite these differences, the insulating LCNO sample has a very similar spectrum to lightly doped LSCO, including the loss of the sharp 2.2 eV peak.\\

At \textbf{q}$\approx$(0 0), the spectral features in the 2-5 eV range are for the most part unchanged for $0.07\leq x \leq0.25$, as shown in Fig.~\ref{fig:widerange}(a).  We notice that between $x$=0.07 and $x$=0.10, there is only a small increase in the low-energy ($\hbar\omega\leq$2 eV) intensity, but then from $x$=0.10 and $x$=0.12 (blue circles) this low energy intensity increases significantly (which can be seen upon inspection of the inset).  A subsequent measurement of the $x$=0.12 sample using the position-sensitive detector with better resolution is also shown on the graph as orange circles with black outline, and overlaps with the original data quite well, confirming that the observed increase was not due to poor energy resolution.  For the heavily doped $x$=0.30 sample, the intensity between 3 eV and 4 eV markedly decreases, resulting in a spectral weight shift towards higher energies.  However, the high-energy tail above 4.5 eV of all of the spectra are similar, with no systematic trend observed.  At the highest doping, $x$=0.35, there is a single high energy peak at $\sim$4.5 eV.\\

%%%%%%%%%%%%%%%%%%%%%%%%%%%%%%%%%%%%%%%%%%%%%%%%%%%%%%%%%%%%%%%%%%%%%%%%%
%==============================================================================
\begin{figure}
\begin{center}
\epsfig{file=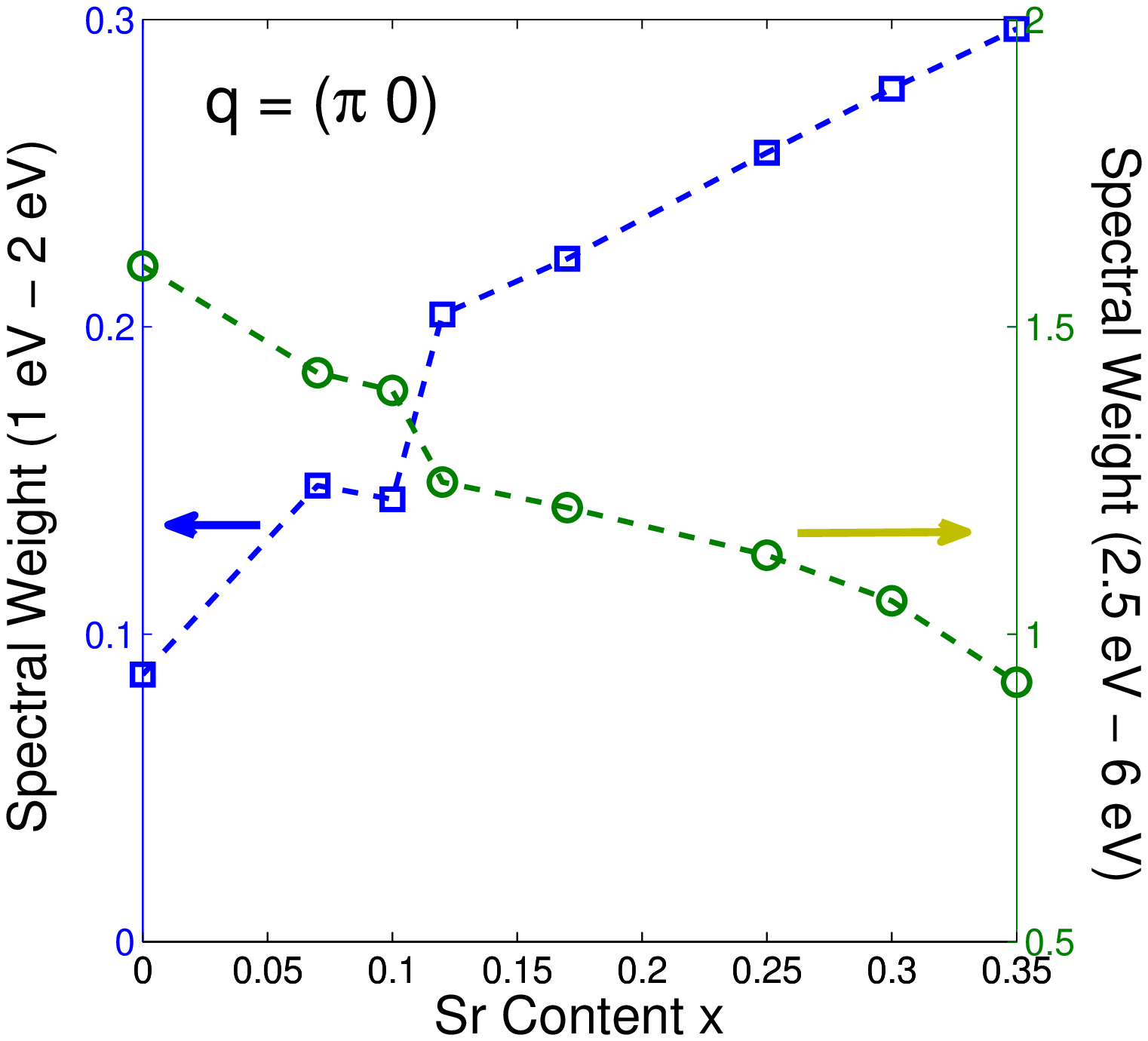,scale=0.35}
\end{center}
\caption{(Color Online) The spectral weight for \textbf{q}=($\pi$, 0) as a function of doping, where the RIXS intensity has been integrated in the low-energy (1 eV -2 eV) and high-energy ranges (2.5 eV - 6 eV).}\label{fig:SpectralWeight}
\end{figure}
%%%%%%%%%%%%%%%%%%%%%%%%%%%%%%%%%%%%%%%%%%%%%%%%%%%%%%%%%%%%%%%%%%%%%%%%%%%

The doping dependence at \textbf{q}=($\pi$,0), shown in Fig.~\ref{fig:widerange}(b) is quite similar to that of \textbf{q}=(0 0).  The prominent peak at $\sim$3.3 eV is already suppressed at $x$=0.07, and little change is seen between $x$=0.07 and $x$=0.10; but a relatively large change in the spectral weight below 2 eV is observed from $x$=0.10 to $x$=0.12.  Like \textbf{q}$\sim$(0 0) at the highest doping, the remnant high-energy feature is a single 4.5 eV peak.  A difference from \textbf{q}=(0,0) is that the spectral weight shift seems to occur monotonically, with the low-energy intensity increasing with doping throughout most of the doping range.  This is shown in Fig.~\ref{fig:SpectralWeight}, in which integrated spectral weights in the low (1-2 eV) and high (2.5-6 eV) energy ranges are compared.  A kink is seen around $x\sim$0.1, above which the low energy spectral weight increases nearly linearly with doping.  In addition, a $\sim$1.5 eV peak is seen for both underdoped and overdoped samples.  Finally, all spectra appear to converge in a narrow energy region around 2.3 eV, suggesting an isosbestic point.  An isobestic point appears at the other \textbf{q} positions as well, and implies a superposition of two spectra whose relative weights are controlled by doping.  From a recent theoretical study by Eckstein et. al.,\cite{Eckstein07} an isobestic point would further imply the existence of sum rules for these RIXS spectra, but due to the lack of data below 1 eV, we are unable to verify this using Fig.~\ref{fig:SpectralWeight}, for example.\\

%==============================================================================
\begin{figure}
\begin{center}
\epsfig{file=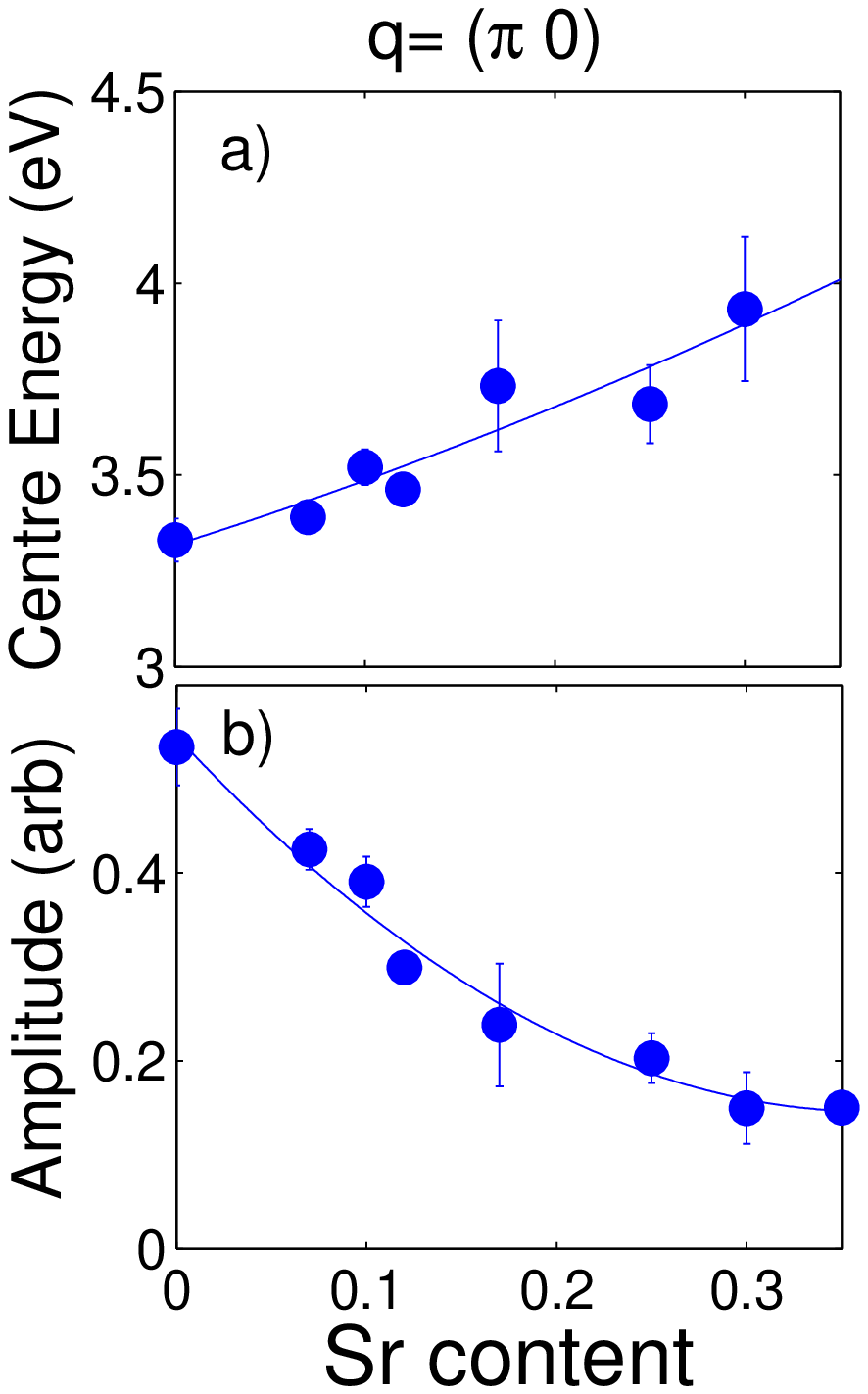,width=4.2in,keepaspectratio,bbllx=17,bblly=220,bburx=593,bbury=610,clip=}
\end{center}
\caption{(Color Online) The (a) center energy and (b) peak intensity of the 3.3 eV peak at \textbf{q}=($\pi$, 0), obtained from the fitting procedure described in the text.  The lines are guides for the eye.}\label{fig:3eV_fitparams}
\end{figure}
%%%%%%%%%%%%%%%%%%%%%%%%%%%%%%%%%%%%%%%%%%%%%%%%%%%%%%%%%%%%%%%%%%%%%%%%%%%

The spectral features at \textbf{q}=($\pi$, 0) above 2 eV were fit to multiple Lorentzian peaks.  Two Lorentzians, with the parameters of the higher energy Lorentzian held to be the same for all dopings, plus a constant background, fit the data reasonably well.  Some of these fits are shown in Fig.~\ref{fig:widerange}(b) as solid lines.  The peak position and intensity parameters of the first peak are plotted in Fig.~\ref{fig:3eV_fitparams}.  This first peak has an energy of 3.3 eV at $x$=0, loses more than half of its intensity and appears to shift $\sim$600 meV to higher energy in the heavily overdoped region.  The other peak, at 4.5 eV  is relatively insensitive to doping and was kept fixed in our fitting.  We note that at the highest doping of $x$=0.35, a high energy intensity at 5.6 eV shows up, but the 4.5 eV peak remains as the predominant feature.\\

As seen in Fig.~\ref{fig:widerange}(c), the \textbf{q}=($\pi$, $\pi$) data features one main peak at high energy.  In contrast to the other \textbf{q} positions, the main high-energy peak appears to be insensitive to doping up to $x$=0.10.  Between $x$=0.10 and $x$=0.25, the peak intensity starts to decrease and continues to do so in the heavily overdoped region.  The high-energy tail remains common for all dopings, and there is a single remnant peak at 5.2 eV for the highly overdoped sample.  Like the other \textbf{q} positions, the low energy spectral weight increases with increased doping.  Note that the momentum dependence of the RIXS spectra, especially along the ($\pi$, 0) direction, is very small for the highly overdoped sample, which exhibits a broad single peak located around 4.5 eV for all momenta.\\

\subsection{Near-IR Feature}
\label{subsect:Near-IR Feature}

\begin{figure}
\begin{center}
\epsfig{file=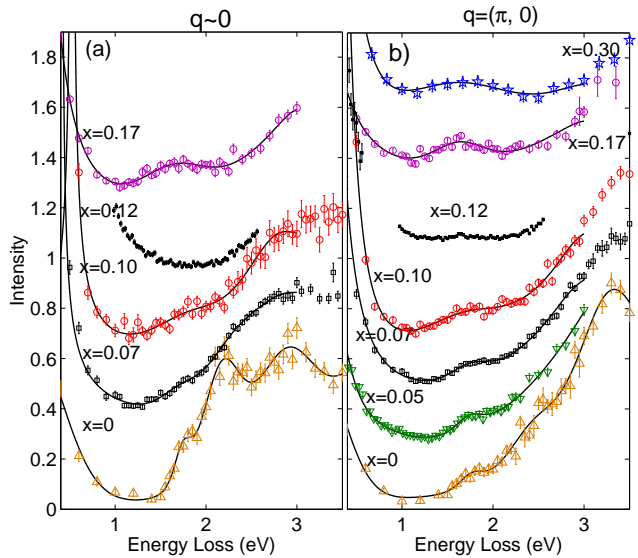,width=4.0in, keepaspectratio}
\end{center}
\caption{(Color Online) The low energy part of the RIXS spectra for momentum transfers of (a) q$\leq$0.1$\pi$ and (b) q=($\pi$,0).  The spectra are displaced along the $y$-axis for clarity, and the solid lines are Lorentzian fits.}
\label{fig:1p7eV_fits}
\end{figure}

Most of the \textbf{q}=(0,0) and \textbf{q}=($\pi$,0) spectra in Fig.~\ref{fig:widerange} exhibit a peak between 1 eV and 2 eV. Fig.~\ref{fig:1p7eV_fits} expands the low energy region, showing the high-resolution spectra.  In Ref~\onlinecite{Ellis08}, a dispersionless feature at around~1.7 eV -1.8 eV in the undoped sample was observed, and this is also seen in Fig.~\ref{fig:1p7eV_fits}(a).  It is even more apparent for the \textbf{q}=($\pi$, 0) data shown in Fig.~\ref{fig:1p7eV_fits}(b). As $x$ increases from zero, this feature at \textbf{q}=($\pi$,0) does not change much for $x$~$\leq$0.10, but appears to vanish for \emph{x}=0.12. It is possible that the peak may be lost in the large continuum intensity which increases dramatically at this doping.  At higher dopings still, a broad peak emerges for $x\geq$0.17 at around the same energy.  It is seen prominently at both \textbf{q}=($\pi$,0) and \textbf{q}=(0,0).\\

%==============================================================================
%==============================================================================

%==============================================================================
\begin{figure}
\begin{center}
\epsfig{file=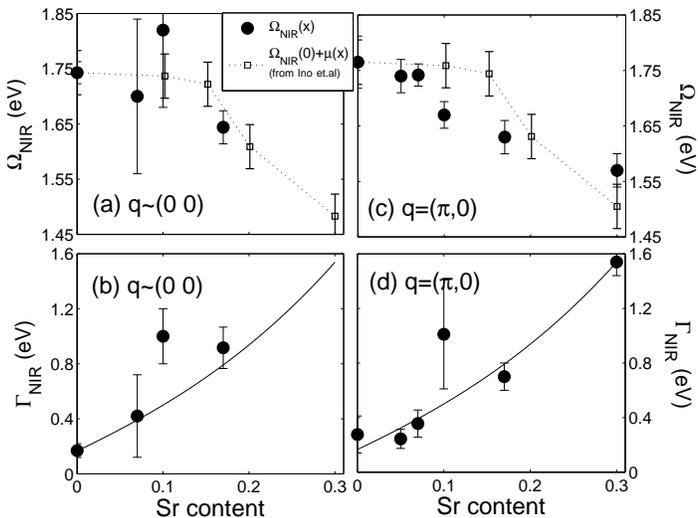,scale=0.45}
\end{center}
\caption{(Color Online) (filled circles) Fit parameters of near-infrared excitation peak vs.doping for
momentum transfers of q$\sim$0 (left) and q=($\pi$,0) (right).  The
parameters are center energy of the NIR peak (top), and full-width-at-half-maximum (bottom).   See text for details of the fitting procedure.  Also shown in (a) and (c) is a curve of the undoped center energy, shifted by the chemical potential obtained from photoemission measurements of Ino et al. \cite{Ino97} (unfilled squares).   The line in graphs (b) and (d) is a guide for the eye.} \label{fig:1p7eV_fit_params}
\end{figure}
%==============================================================================

Fits to the RIXS spectra are shown in Fig.~\ref{fig:1p7eV_fits} as the solid lines. To account for the different energy resolution used in our measurements, the intrinsic cross-section (Lorentzian) was convolved with the instrumental resolution (sum of Gaussian and Lorentzian to the fourth power) in our fitting.  The resultant fit parameters of this NIR peak, namely the center energy $\rm \Omega_{NIR}$ and the full-width-at-half-maximum $\rm \Gamma_{NIR}$, are plotted as the filled circles in Fig.~\ref{fig:1p7eV_fit_params}.  The heavily overdoped samples were not fit, except for $x$=0.30 at \textbf{q}=($\pi$,0),  because the resolution and/or number of data points was not high enough to yield reliable fit parameters.  From inspection of Fig.~\ref{fig:widerange}(b), the peak still seems to exist up to $x$=0.35.  The center energy is roughly constant for x$\leq$0.07, from which it monotonically decreases by $\sim$150 meV as $x$ increases up to 0.30.  The width, however, changes dramatically; for \emph{x}=0.17 it is approximately double that for x$<$0.10, and doubles again for \emph{x}=0.30.  We will discuss the physical origin of this feature in detail below.\\

\subsection{Barium-doped x=1/8 Sample}
\label{subsect:Ba doped}
\begin{figure}
\begin{center}
\epsfig{file=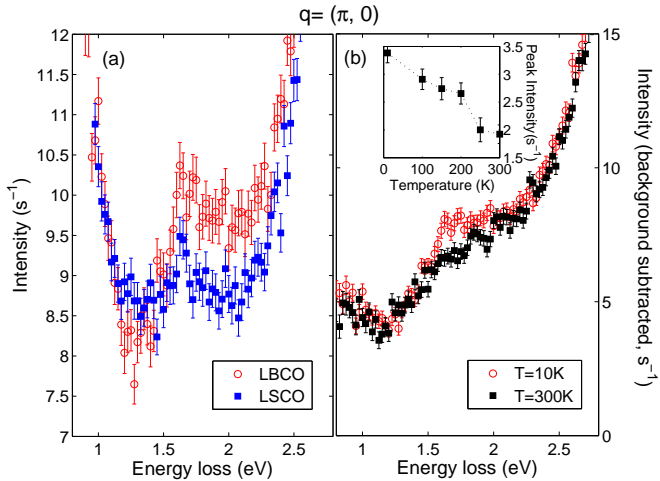,width=3.6in, keepaspectratio}
\end{center}
\caption{(Color Online) (a) Comparison of $\rm La_{1.88}Sr_{0.12}CuO_4$ and $\rm La_{1.875}Ba_{0.125}CuO_4$ spectra measured at \textbf{q}=($\pi$,0) and T=28 K.  (b) Temperature dependence of the $\rm La_{1.875}Ba_{0.125}CuO_4$ \textbf{q}=($\pi$,0) spectra, after subtraction of elastic line.  The inset shows the intensity of the NIR component as obtained from fits described in the text. } \label{fig:compare_LSCO_LBCO}
\end{figure}

A comparison of the low energy part of the LSCO \emph{x}=0.12 and LBCO $x$=0.125 spectra, measured under identical conditions, is shown in Fig.~\ref{fig:compare_LSCO_LBCO}(a).  Data are normalized such that the intensities at 10 eV are equal (there is generally good agreement in the high energy spectra up to 10 eV, not shown). Fig.~\ref{fig:compare_LSCO_LBCO}(a) shows good overlap of the spectra up to 1.2 eV, after which the near-infrared peak begins to emerge, which is at least twice as intense in LBCO.  The overlap of the spectra, aside from the near-infrared peak, suggests that most of the features  seen in the spectra arise purely from holes doped into the copper-oxygen plane.  They are not (aside from the NIR peak) affected by whether the dopant is Sr and Ba, and are independent of the fact that, at the low temperature, the LBCO sample is structurally different from the LSCO sample.  The relative enhancement of the NIR peak in LBCO could be due to either structural or electronic differences. That is, LBCO has a low-temperature tetragonal structure and has static charge stripe order \cite{Kim08}, while LSCO has a low-temperature orthorhombic structure, without static charge order.  It should be noted that we did not observe an obvious enhancement of the NIR peak at the stripe-ordering wave-vector; the peak was still observed at \textbf{q}=($\pi$/2,0), but the background was higher (our measurements of momentum dependence will be presented elsewhere, but our low-energy elastic tails were too high to verify the result of Ref.~\onlinecite{Wakimoto09}).\\

The temperature dependence is summarized in Fig.~\ref{fig:compare_LSCO_LBCO}(b) and its inset.  Six temperatures were measured.  Changes in the NIR peak were very gradual and we only show the highest and lowest temperatures in the figure for clarity.  The quasi-elastic background, which is expected to depend on temperature, was subtracted by using a fit to the energy-gain ($\hbar\omega <$0) side of the spectrum as the background.\cite{Ellis10}  That the spectra overlap with the exception of the NIR feature indicates that the background subtraction procedure is correct, and that the only temperature dependent part is the NIR feature.   The upturn of the intensity at $\sim$1.2 eV is consistent with that observed in Ref.~\onlinecite{Wakimoto09}.  While the absolute intensity at 1.7 eV in Fig.~\ref{fig:compare_LSCO_LBCO}(b) changed only by $\sim$15\%, the spectra above and below this feature are virtually identical at all temperatures, which suggests that (a) the change at 1.7 eV is not due to sample moving or other experimental artifact; and (b) the 1.7 eV peak is particularly sensitive to temperature at this \textbf{q}-position.\\

To fit the NIR peak, the background-subtracted spectrum in the energy range from 0.5 eV to 3 eV was modeled as a Lorentzian peak, plus a temperature independent background.  The peak position shifts slightly downward by 40$\pm$15 meV, while the width was constant to within error (exception for T=300 K).  The inset of Fig.~\ref{fig:compare_LSCO_LBCO}(b) shows the peak intensity decreasing continuously by nearly a factor of two over this temperature range.\\

%==============================================================================
%==============================================================================

\section{Discussion}
\label{sect: Discussion}
\vspace{-3 mm}

%%%%%%%%%%%%%%%%%%%%%%%%%%%%%%%%%%%%%%%%%%%%%%%%%%%%%%%%%%%%%%%%%%%%%%%%%%%
%%%%%%%%%%%%%%%%%%%%%%%%%%%%%%%%%%%%%%%%%%%%%%%%%%%%%%%%%%%%%%%%%%%%%%%%%%%%%%%%%%%%%%%%%%
\begin{figure}
\begin{center}
\includegraphics[width=3.3in]{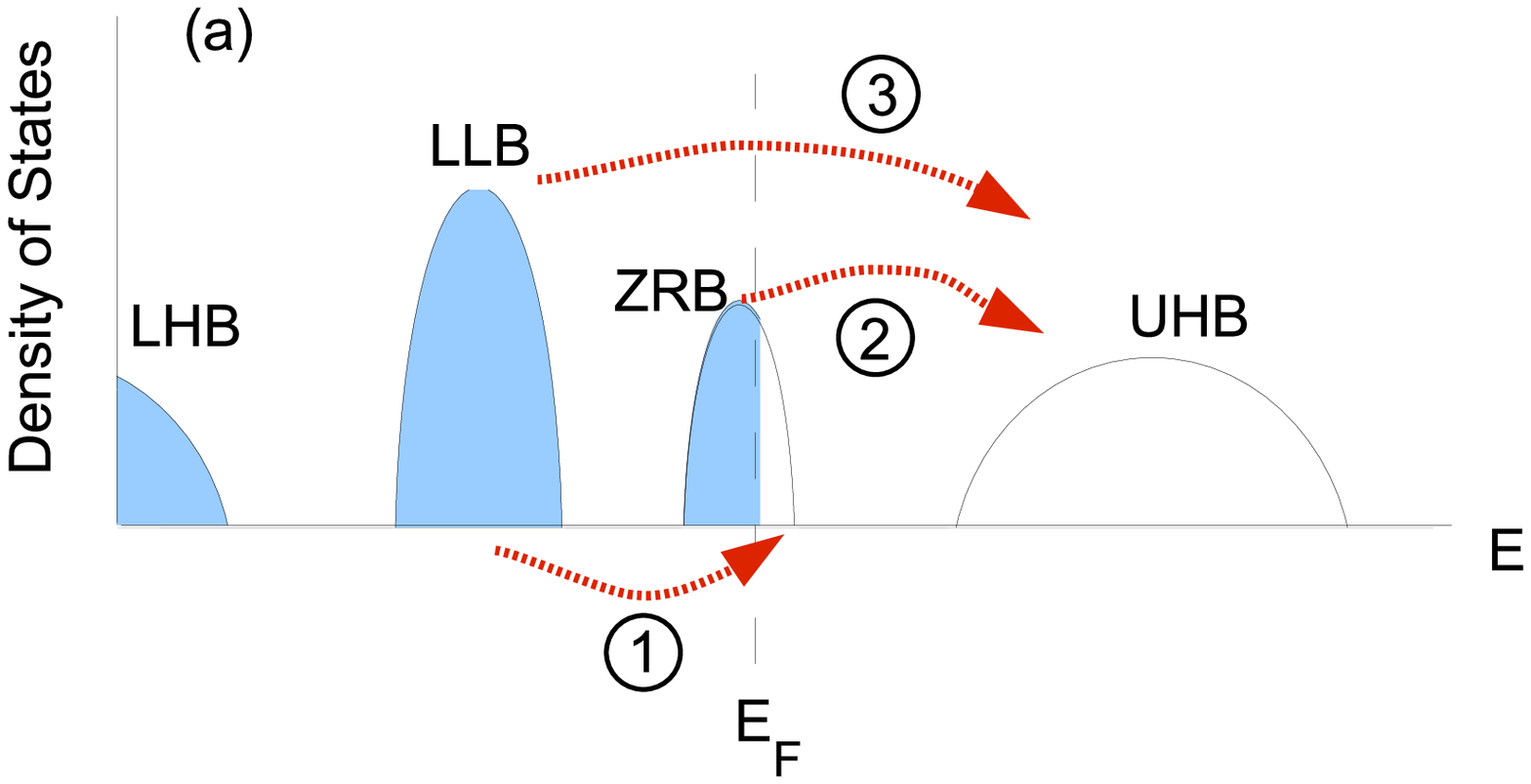} \\
\includegraphics[width=3.3in]{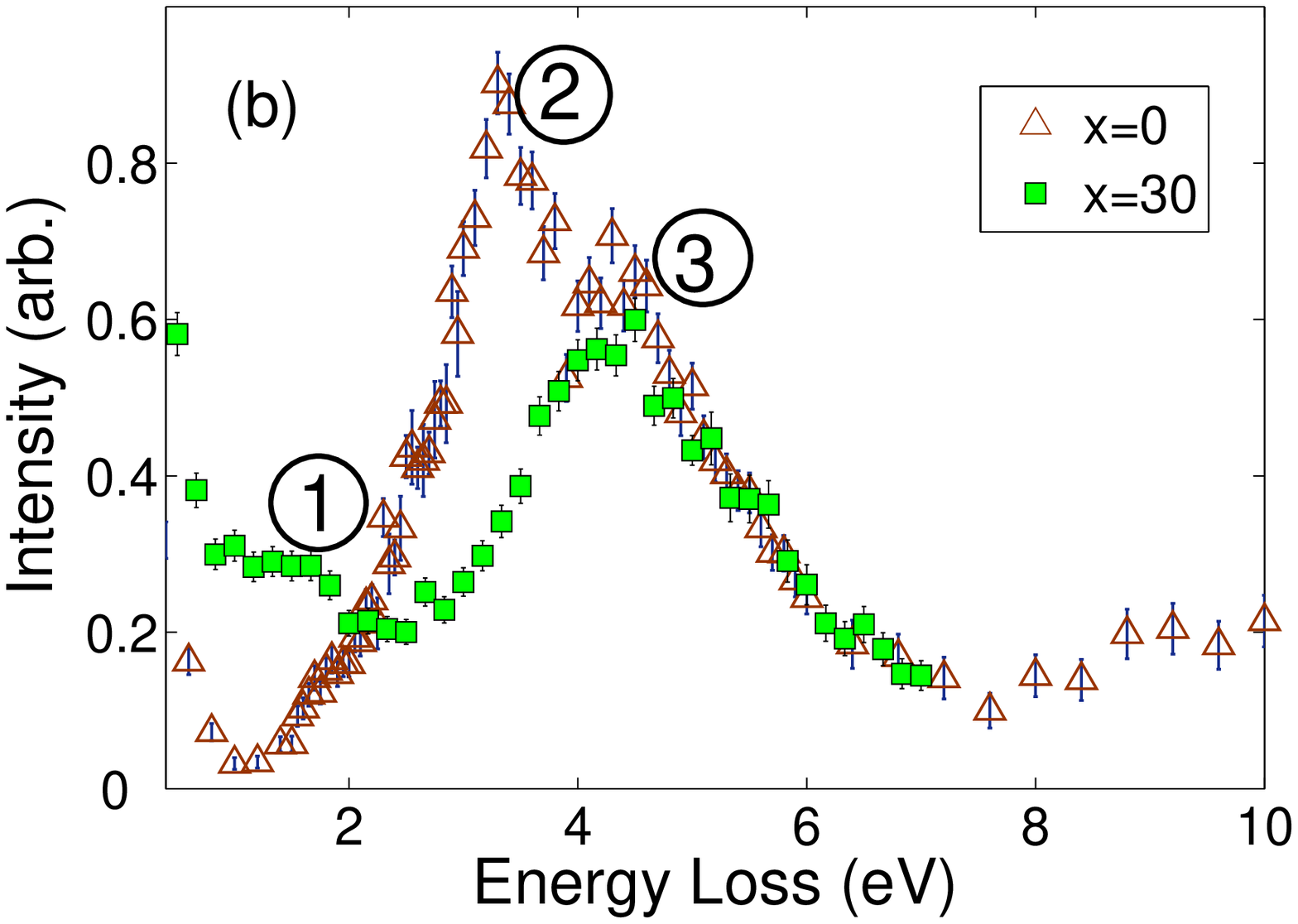}
\end{center}
\caption{(Color Online) (a) Rigid band model density of states vs. energy profile, showing the lower-lying bands (LLB), Zhang-Rice band (ZRB) and the upper Hubbard band (UHB).  Grey filling represents the occupied states.  The lower Hubbard band (LHB) is also shown for completeness.  Transitions between these bands, which we assign to peaks in the RIXS spectra and signified by the arrows, are labeled by number : LLB$\rightarrow$ ZRB (1), ZRB$\rightarrow$UHB (2), LLB$\rightarrow$UHB (3).  The corresponding transitions in the spectra are shown in (b), for $x$=0 and $x$=0.30 at \textbf{q}=($\pi$, 0).  For $x$=0, $E_F$, as shown by the dashed line in (a), would be shifted to the right past the edge of the ZRB, while $x$=0.30, would correspond to $E_F$ being near the left edge of the ZRB.} \label{fig:bandpeaka}
\end{figure}
%%%%%%%%%%%%%%%%%%%%%%%%%%%%%%%%%%%%%%%%%%%%%%%%%%%%%%%%%%%%%%%%%%%%%%%%%%%%%%%%%%%%%%%%%%%

The following discussion is qualitatively based on the general picture of RIXS as a probe of interband transitions, developed by previous authors,\cite{Tsutsui99,Nomura05,Markiewicz06b} and recently used to explain the momentum dependence of the $\rm Cu_2O$ RIXS spectrum.\cite{Kim10}  Fig.~\ref{fig:bandpeaka} shows a scenario where the peaks in the RIXS spectra are transitions between various bands, which are assumed to be rigid (i.e., doping-independent density of states).  The bands under consideration are Zhang-Rice band (ZRB) states near the Fermi level, the upper Hubbard band (UHB) above the Fermi level, and ``lower-lying" bands (LLB) which we define as more than 1 eV below the Fermi level, but still relatively near compared to the lower Hubbard band.  Thus it includes copper $d$ states (other than $d_{x^2-y^2}$) and non-bonding oxygen states.  Exact bands are not assigned, due to uncertainties in models and parameters; instead this discussion is limited to general energy regions, near where it is generally accepted that bands exist.  A schematic sketch of the density of states according to this picture is shown in Fig.~\ref{fig:bandpeaka}(a), in which transitions are labeled as ``1", ``2" and ``3".  In the \textbf{q}=($\pi$, 0) spectra, we assign the three main peaks as shown in Fig.~\ref{fig:bandpeaka}(b).  As hole doping increases, the Fermi level shifts downward, traversing the ZRB, which increases the spectral weight of transition 1, while decreasing the transition 2 spectral weight.  In addition, there will be intraband transitions within the ZRB, which contribute to the Drude-like intensity at lower energies.\\

The transition between LLB and UHB, which are both far from the Fermi level, would therefore not be directly affected by the Fermi level shifting.  From our spectra, it naturally follows that the doping-independent 4.5 eV peak would be a candidate for this transition.  The energy is also reasonable; for example Nomura and Igarashi calculated that a $\sim$5 eV peak should arise from transitions to the UHB from mixed states a few eV below the Fermi level.\cite{Nomura05} \\

The transition between LLB and ZRB could be the NIR peak.  The energy of $\sim$1.5 eV is within reason for transitions from lower-lying states to the ZRB near the Fermi level (e.g. the ``1 eV peaks'' observed in ARPES of various other cuprates,\cite{Tobin92,Olson95,Pothuizen97,Durr00} and the bands calculated by Markiewicz and Bansil\cite{Markiewicz06b} or Wagner et.~al.\cite{Wagner91}, for example)  As the Fermi level moves to lower energy as hole doping is increased, more empty states in the ZRB become available for such an excitation; the intensity will increase, and the energy of the transition will shift downwards.  In Fig.~\ref{fig:1p7eV_fit_params} we plot the Fermi level decrease superimposed with the peak energy, showing that they are comparable, and supporting the interpretation of the NIR feature as excitation(s) between the LLB and the ZRB. The physical nature of this excitation of course depends on the specific characteristics of the LLB. If one considers Cu $d$-states other than $d_{x^2-y^2}$ orbital, excitation ``1" takes on the character of a crystal field excitation, which is typically localized and therefore is relatively sharp and dispersionless, and was our previous interpretation.\cite{Ellis08}  Excitations involving non-bonding oxygen states would be in principle similarly localized, and exhibit similarly sharp features.  The NIR feature in the undoped sample does exhibit such localized properties, but the dramatic increase of the NIR peak width in the overdoped regime, shown in Fig.~\ref{fig:1p7eV_fits}, suggests that the lifetime of this excitation becomes very short. \\

It is tempting to associate some of the LLB with stripes, given the enhancement of the relative intensity of the NIR peak in LBCO compared to LSCO shown in Fig.~\ref{fig:compare_LSCO_LBCO}. In fact, recent calculations do associate some of the bands close to the Fermi level with stripe ordering,\cite{Fleck01} in particular the wide band centered at \textbf{k}=($\pi$,0). If  ``1" is then considered as a stripe-related excitation, it would be expected to broaden as hole doping increases, since the stripes in real space become more disordered and fluctuating.\\

The last transition considered in Fig.~\ref{fig:bandpeaka} is from the ZRB to the UHB.  The peak's energy of 3.3 eV is roughly equal to the difference between the other two peak energies ($\sim$1.5 eV and 4.5 eV).  Its intensity would decrease and shift to higher energies as the ZRB is depleted of occupied states due to the shifting Fermi level.  This effect was discussed in the theoretical RIXS studies of Tsutsui et al.\cite{Tsutsui99} and Markiewicz and Bansil\cite{Markiewicz06b}, and earlier studies by Wagner et.~al.\cite{Wagner91}  That more than half of the intensity disappears by high doping (Fig.~\ref{fig:3eV_fitparams}(b)) suggests that the Fermi level traversed the greater part of the ZRB.  This would be in partial agreement with the recent near-edge absorption study of Peets et al.\cite{Peets09}, who observed that the oxygen \emph{K}-edge absorption peak grows with doping but then saturates above $x$=0.22, a possible indication that there are no additional states to be gained from the ZRB as the Fermi level passes its bottom.\\

However, an energy shift of $\sim$0.6 eV seen in Fig.~\ref{fig:3eV_fitparams}(a) seems inconsistent with the known Fermi level shift of 0.3 eV.\cite{Ino97}  One would expect it to shift \emph{less} than the Fermi level, yet here it shifts by more than twice as much.  It is quite possible that part of this major discrepancy is due to the specific model used in the fitting.  When an additional high-energy peak at 5.6 eV was included to better fit the data, the energy shift up to $x$=0.25 seemed to correspond more closely to the Fermi-level shift.  Such an additional peak might be a lower Hubbard band (LHB) to ZRB transition.  On the other hand, this discrepancy may be also due to a fundamental failure of the rigid band model assumption.  In the recent dynamic mean field study by de'Medici et al.\cite{Medici09}, for example, the lower-energy part of the UHB is greatly reduced upon hole doping (less than 20\% at $x$=0.15), which would result in increased transition energy.  While this could qualitatively explain a larger upward shift in the peak as the doping increases, theoretical calculations will be needed to explore these questions rigourously.\\

Finally, we briefly discuss the 2.2 eV peak at \textbf{q}$\sim$(0 0), which sharply decreases in intensity upon light doping, whether by Sr or Ni, as shown in Fig.~\ref{fig:lcno}.  It is possibly an excitonic excitation,\cite{Zhang98} although such an interpretation has been contested in the literature.\cite{Takahashi09}  In their recent electron energy-loss spectroscopy study of $\rm Ca_{2-x}Na_xCuO_2Cl_2$, Schuster et al. explained the decrease of the exciton peak on doping as an unbinding of the exciton due to screening from mobile charge carriers.\cite{Schuster09}  However, the doped holes in LCNO are not mobile, although they do reside in the Cu-O plane,\cite{Hiraka09} which therefore casts doubt on this picture for the 2.2 eV peak reduction in the case of LCNO.   We also note that the LCNO sample is magnetically ordered, while the doped LSCO sample is not, so it would also seem that the loss of magnetic order does not play a role.  A more detailed study of a series of lightly doped samples may clarify the behavior of this peak.\\

\section{Conclusions}
\label{sect: Conclusion}
In summary, we have carried out an investigation of RIXS spectra in a series of doped lanthanum cuprates as a function of doping and momentum transfer, with a view of prompting more detailed theoretical studies.  A 2.2 eV peak observed at the zone center rapidly diminishes, both with Sr and Ni doping.  Further mesaurements in the low doping region may help to explain this sharp reduction.  In the zone-boundary \textbf{q}=($\pi$, 0) spectra, we were able to characterize three major spectral features, with spectral weight being transferred systematically from the high energy to the low energy peaks as doping increases.  At higher energies, the spectra can be described as two peaks over a wide doping range, one at 4.5 eV which is independent of doping, the other at 3.3 eV which decreases in intensity, and increases in energy approximately linearly with doping.  We assigned these peaks to transitions from lower-lying bands (LLB) $\rightarrow$ upper-Hubbard band (UHB) and Zhang-Rice band (ZRB) $\rightarrow$ upper-Hubbard band (UHB) transitions respectively.  At lower energies a near-infrared (NIR) peak was observed, we assign this to a LLB$\rightarrow$ZRB transition.  It has zero dispersion along the (0 0)-($\pi$, 0) direction.  A peak appears at around the same $\sim$1.7 eV energy for all dopings, although around $x$=0.12 it appears small compared to the continuum intensity which increases at this doping.  In the overdoped region, the NIR peak width increases dramatically, and the peak energy red-shifts with increased doping.  Our comparative study of LSCO and LBCO suggests a possible connection between the NIR peak and charge stripes.\\

\section{Acknowledgements}
\label{sect: Acknowledgements}
We would like to thank Suichi Wakimoto for sharing his data from Ref.~\onlinecite{Wakimoto05}.
The work at University of
Toronto was supported by Natural Sciences and Engineering Research
Council of Canada. The work at Brookhaven was supported by the U. S.
DOE, Office of Science Contract No. DE-AC02-98CH10886. Use of the
Advanced Photon Source was supported by the U. S. DOE, Office of
Science, Office of Basic Energy Sciences, under Contract No.
W-31-109-ENG-38.


\begin{thebibliography}{56}
\expandafter\ifx\csname natexlab\endcsname\relax\def\natexlab#1{#1}\fi
\expandafter\ifx\csname bibnamefont\endcsname\relax
  \def\bibnamefont#1{#1}\fi
\expandafter\ifx\csname bibfnamefont\endcsname\relax
  \def\bibfnamefont#1{#1}\fi
\expandafter\ifx\csname citenamefont\endcsname\relax
  \def\citenamefont#1{#1}\fi
\expandafter\ifx\csname url\endcsname\relax
  \def\url#1{\texttt{#1}}\fi
\expandafter\ifx\csname urlprefix\endcsname\relax\def\urlprefix{URL }\fi
\providecommand{\bibinfo}[2]{#2}
\providecommand{\eprint}[2][]{\url{#2}}

\bibitem[{\citenamefont{Bednorz and Muller}(1988)}]{Bednorz88}
\bibinfo{author}{\bibfnamefont{J.~G.} \bibnamefont{Bednorz}} \bibnamefont{and}
  \bibinfo{author}{\bibfnamefont{K.~A.} \bibnamefont{Muller}},
  \bibinfo{journal}{Revews of Modern Physics} \textbf{\bibinfo{volume}{60}},
  \bibinfo{pages}{585} (\bibinfo{year}{1988}).

\bibitem[{\citenamefont{Tranquada et~al.}(1995)\citenamefont{Tranquada,
  Sternlieb, Axe, Nakumura, and Uchida}}]{Tranquada95}
\bibinfo{author}{\bibfnamefont{J.~M.} \bibnamefont{Tranquada}},
  \bibinfo{author}{\bibfnamefont{B.~J.} \bibnamefont{Sternlieb}},
  \bibinfo{author}{\bibfnamefont{J.~D.} \bibnamefont{Axe}},
  \bibinfo{author}{\bibfnamefont{Y.}~\bibnamefont{Nakumura}}, \bibnamefont{and}
  \bibinfo{author}{\bibfnamefont{S.}~\bibnamefont{Uchida}},
  \bibinfo{journal}{Nature} \textbf{\bibinfo{volume}{375}},
  \bibinfo{pages}{561} (\bibinfo{year}{1995}).

\bibitem[{\citenamefont{Zhang and Rice}(1988)}]{Zhang88}
\bibinfo{author}{\bibfnamefont{F.~C.} \bibnamefont{Zhang}} \bibnamefont{and}
  \bibinfo{author}{\bibfnamefont{T.~M.} \bibnamefont{Rice}},
  \bibinfo{journal}{\prb} \textbf{\bibinfo{volume}{37}}, \bibinfo{pages}{3759}
  (\bibinfo{year}{1988}).

\bibitem[{\citenamefont{Ino et~al.}(2002)\citenamefont{Ino, Kim, Nakamura,
  Yoshida, Mizokawa, Fujimori, Shen, Kakeshita, Eisaki, and Uchida}}]{Ino02}
\bibinfo{author}{\bibfnamefont{A.}~\bibnamefont{Ino}},
  \bibinfo{author}{\bibfnamefont{C.}~\bibnamefont{Kim}},
  \bibinfo{author}{\bibfnamefont{M.}~\bibnamefont{Nakamura}},
  \bibinfo{author}{\bibfnamefont{T.}~\bibnamefont{Yoshida}},
  \bibinfo{author}{\bibfnamefont{T.}~\bibnamefont{Mizokawa}},
  \bibinfo{author}{\bibfnamefont{A.}~\bibnamefont{Fujimori}},
  \bibinfo{author}{\bibfnamefont{Z.~X.} \bibnamefont{Shen}},
  \bibinfo{author}{\bibfnamefont{T.}~\bibnamefont{Kakeshita}},
  \bibinfo{author}{\bibfnamefont{H.}~\bibnamefont{Eisaki}}, \bibnamefont{and}
  \bibinfo{author}{\bibfnamefont{S.}~\bibnamefont{Uchida}},
  \bibinfo{journal}{\prb} \textbf{\bibinfo{volume}{65}},
  \bibinfo{pages}{094504} (\bibinfo{year}{2002}).

\bibitem[{\citenamefont{Yoshida et~al.}(2003)\citenamefont{Yoshida, Zhou,
  Sasagawa, Yang, Bogdanov, Lanzara, Hussain, Mizokawa, Fujimori, Eisaki,Shen,Kakeshita,Uchida}}]{Yoshida03}
\bibinfo{author}{\bibfnamefont{T.}~\bibnamefont{Yoshida}},
  \bibinfo{author}{\bibfnamefont{X.~J.} \bibnamefont{Zhou}},
  \bibinfo{author}{\bibfnamefont{T.}~\bibnamefont{Sasagawa}},
  \bibinfo{author}{\bibfnamefont{W.~L.} \bibnamefont{Yang}},
  \bibinfo{author}{\bibfnamefont{P.~V.} \bibnamefont{Bogdanov}},
  \bibinfo{author}{\bibfnamefont{A.}~\bibnamefont{Lanzara}},
  \bibinfo{author}{\bibfnamefont{Z.}~\bibnamefont{Hussain}},
  \bibinfo{author}{\bibfnamefont{T.}~\bibnamefont{Mizokawa}},
  \bibinfo{author}{\bibfnamefont{A.}~\bibnamefont{Fujimori}},
  \bibinfo{author}{\bibfnamefont{H.}~\bibnamefont{Eisaki}},
  \bibinfo{author}{\bibfnamefont{Z.-X.}~\bibnamefont{Shen}},
  \bibinfo{author}{\bibfnamefont{T.}~\bibnamefont{Kakeshita}},
  \bibinfo{author}{\bibfnamefont{S.}~\bibnamefont{Uchida}},
  \bibinfo{journal}{\prl} \textbf{\bibinfo{volume}{91}},
  \bibinfo{pages}{027001} (\bibinfo{year}{2003}).

\bibitem[{\citenamefont{Yoshida et~al.}(2006)\citenamefont{Yoshida, Zhou,
  Tanaka, Yang, Hussain, Shen, Fujimori, Sahrakorpi, Lindroos, Markiewicz
  et~al.}}]{Yoshida06}
\bibinfo{author}{\bibfnamefont{T.}~\bibnamefont{Yoshida}},
  \bibinfo{author}{\bibfnamefont{X.~J.} \bibnamefont{Zhou}},
  \bibinfo{author}{\bibfnamefont{K.}~\bibnamefont{Tanaka}},
  \bibinfo{author}{\bibfnamefont{W.~L.} \bibnamefont{Yang}},
  \bibinfo{author}{\bibfnamefont{Z.}~\bibnamefont{Hussain}},
  \bibinfo{author}{\bibfnamefont{Z.~X.} \bibnamefont{Shen}},
  \bibinfo{author}{\bibfnamefont{A.}~\bibnamefont{Fujimori}},
  \bibinfo{author}{\bibfnamefont{S.}~\bibnamefont{Sahrakorpi}},
  \bibinfo{author}{\bibfnamefont{M.}~\bibnamefont{Lindroos}},
  \bibinfo{author}{\bibfnamefont{R.~S.} \bibnamefont{Markiewicz}},
  \bibnamefont{et~al.}, \bibinfo{journal}{\prb} \textbf{\bibinfo{volume}{74}},
  \bibinfo{pages}{224510} (\bibinfo{year}{2006}).

\bibitem[{\citenamefont{Doiron-Leyraud
  et~al.}(2007)\citenamefont{Doiron-Leyraud, Proust, LeBoeuf, Levallois,
  Bonnemaison, Liang, Bonn, Hardy, and Taillefer}}]{DorionLeyraud07}
\bibinfo{author}{\bibfnamefont{N.}~\bibnamefont{Doiron-Leyraud}},
  \bibinfo{author}{\bibfnamefont{C.}~\bibnamefont{Proust}},
  \bibinfo{author}{\bibfnamefont{D.}~\bibnamefont{LeBoeuf}},
  \bibinfo{author}{\bibfnamefont{J.}~\bibnamefont{Levallois}},
  \bibinfo{author}{\bibfnamefont{J.-B.} \bibnamefont{Bonnemaison}},
  \bibinfo{author}{\bibfnamefont{R.}~\bibnamefont{Liang}},
  \bibinfo{author}{\bibfnamefont{D.~A.} \bibnamefont{Bonn}},
  \bibinfo{author}{\bibfnamefont{W.~N.} \bibnamefont{Hardy}}, \bibnamefont{and}
  \bibinfo{author}{\bibfnamefont{L.}~\bibnamefont{Taillefer}},
  \bibinfo{journal}{Nature} \textbf{\bibinfo{volume}{447}},
  \bibinfo{pages}{565} (\bibinfo{year}{2007}).

\bibitem[{\citenamefont{LeBoeuf et~al.}(2007)\citenamefont{LeBoeuf,
  Doiron-Leyraud, Levallois, Daou, Bonnemaison, Hussey, Balicas, Ramshaw,
  Liang, Bonn et~al.}}]{LeBoeuf07}
\bibinfo{author}{\bibfnamefont{D.}~\bibnamefont{LeBoeuf}},
  \bibinfo{author}{\bibfnamefont{N.}~\bibnamefont{Doiron-Leyraud}},
  \bibinfo{author}{\bibfnamefont{J.}~\bibnamefont{Levallois}},
  \bibinfo{author}{\bibfnamefont{R.}~\bibnamefont{Daou}},
  \bibinfo{author}{\bibfnamefont{J.-B.} \bibnamefont{Bonnemaison}},
  \bibinfo{author}{\bibfnamefont{N.~E.} \bibnamefont{Hussey}},
  \bibinfo{author}{\bibfnamefont{L.}~\bibnamefont{Balicas}},
  \bibinfo{author}{\bibfnamefont{B.~J.} \bibnamefont{Ramshaw}},
  \bibinfo{author}{\bibfnamefont{R.}~\bibnamefont{Liang}},
  \bibinfo{author}{\bibfnamefont{D.~A.} \bibnamefont{Bonn}},
  \bibnamefont{et~al.}, \bibinfo{journal}{Nature}
  \textbf{\bibinfo{volume}{450}}, \bibinfo{pages}{533} (\bibinfo{year}{2007}).

\bibitem[{qua()}]{quantum_osc_note}
\bibinfo{note}{See [J. Phys.: Condens. Matter, \textbf{21}, 164212 (2009)] and
  references therein.}

\bibitem[{\citenamefont{Lyons et~al.}(1988)\citenamefont{Lyons, Fleury,
  Remeika, Cooper, and Negran}}]{Lyons88}
\bibinfo{author}{\bibfnamefont{K.~B.} \bibnamefont{Lyons}},
  \bibinfo{author}{\bibfnamefont{P.~A.} \bibnamefont{Fleury}},
  \bibinfo{author}{\bibfnamefont{J.~P.} \bibnamefont{Remeika}},
  \bibinfo{author}{\bibfnamefont{A.~S.} \bibnamefont{Cooper}},
  \bibnamefont{and} \bibinfo{author}{\bibfnamefont{T.~J.}
  \bibnamefont{Negran}}, \bibinfo{journal}{\prb} \textbf{\bibinfo{volume}{37}},
  \bibinfo{pages}{2353} (\bibinfo{year}{1988}).

\bibitem[{\citenamefont{Sugai et~al.}(1988)\citenamefont{Sugia, Shamoto, and
  Sato}}]{Sugai88}
\bibinfo{author}{\bibfnamefont{S.}~\bibnamefont{Sugai}},
  \bibinfo{author}{\bibfnamefont{S.I.}~\bibnamefont{Shamoto}}, \bibnamefont{and}
  \bibinfo{author}{\bibfnamefont{M.}~\bibnamefont{Sato}},
  \bibinfo{journal}{\prb} \textbf{\bibinfo{volume}{38}}, \bibinfo{pages}{6436}
  (\bibinfo{year}{1988}).

\bibitem[{\citenamefont{Salamon et~al.}(1995)\citenamefont{Salamon, Liu, Klein,
  Karlow, Cooper, Cheong, Lee, and Ginsberg}}]{Salamon95}
\bibinfo{author}{\bibfnamefont{D.}~\bibnamefont{Salamon}},
  \bibinfo{author}{\bibfnamefont{R.}~\bibnamefont{Liu}},
  \bibinfo{author}{\bibfnamefont{M.~V.} \bibnamefont{Klein}},
  \bibinfo{author}{\bibfnamefont{M.~A.} \bibnamefont{Karlow}},
  \bibinfo{author}{\bibfnamefont{S.~L.} \bibnamefont{Cooper}},
  \bibinfo{author}{\bibfnamefont{S.~W.} \bibnamefont{Cheong}},
  \bibinfo{author}{\bibfnamefont{W.C.}~\bibnamefont{Lee}}, \bibnamefont{and}
  \bibinfo{author}{\bibfnamefont{D.~M.} \bibnamefont{Ginsberg}},
  \bibinfo{journal}{\prb} \textbf{\bibinfo{volume}{51}}, \bibinfo{pages}{6617}
  (\bibinfo{year}{1995}).

\bibitem[{\citenamefont{Uchida et~al.}(1991)\citenamefont{Uchida, Ido, Takagi,
  Arima, Tokura, and Tajima}}]{Uchida91}
\bibinfo{author}{\bibfnamefont{S.}~\bibnamefont{Uchida}},
  \bibinfo{author}{\bibfnamefont{T.}~\bibnamefont{Ido}},
  \bibinfo{author}{\bibfnamefont{H.}~\bibnamefont{Takagi}},
  \bibinfo{author}{\bibfnamefont{T.}~\bibnamefont{Arima}},
  \bibinfo{author}{\bibfnamefont{Y.}~\bibnamefont{Tokura}}, \bibnamefont{and}
  \bibinfo{author}{\bibfnamefont{S.}~\bibnamefont{Tajima}},
  \bibinfo{journal}{\prb} \textbf{\bibinfo{volume}{43}}, \bibinfo{pages}{7942}
  (\bibinfo{year}{1991}).

\bibitem[{\citenamefont{Uchida et~al.}(1996)\citenamefont{Uchida, Tamasaku, and
  Tajima}}]{Uchida96}
\bibinfo{author}{\bibfnamefont{S.}~\bibnamefont{Uchida}},
  \bibinfo{author}{\bibfnamefont{K.}~\bibnamefont{Tamasaku}}, \bibnamefont{and}
  \bibinfo{author}{\bibfnamefont{S.}~\bibnamefont{Tajima}},
  \bibinfo{journal}{Phys. Rev. B} \textbf{\bibinfo{volume}{53}},
  \bibinfo{pages}{14558} (\bibinfo{year}{1996}).

\bibitem[{\citenamefont{Kishida et~al.}(2003)\citenamefont{Kishida, Ono, Sawa,
  Kawasaki, Tokura, and Okamoto}}]{Kishida03}
\bibinfo{author}{\bibfnamefont{H.}~\bibnamefont{Kishida}},
  \bibinfo{author}{\bibfnamefont{M.}~\bibnamefont{Ono}},
  \bibinfo{author}{\bibfnamefont{A.}~\bibnamefont{Sawa}},
  \bibinfo{author}{\bibfnamefont{M.}~\bibnamefont{Kawasaki}},
  \bibinfo{author}{\bibfnamefont{Y.}~\bibnamefont{Tokura}}, \bibnamefont{and}
  \bibinfo{author}{\bibfnamefont{H.}~\bibnamefont{Okamoto}},
  \bibinfo{journal}{Phys. Rev. B} \textbf{\bibinfo{volume}{68}},
  \bibinfo{pages}{075101} (\bibinfo{year}{2003}).

\bibitem[{\citenamefont{Fink et~al.}(1994)\citenamefont{Fink, Nucker,
  Pellegrin, Romberg, Alexander, and Knupfer}}]{Fink94}
\bibinfo{author}{\bibfnamefont{J.}~\bibnamefont{Fink}},
  \bibinfo{author}{\bibfnamefont{N.}~\bibnamefont{Nucker}},
  \bibinfo{author}{\bibfnamefont{E.}~\bibnamefont{Pellegrin}},
  \bibinfo{author}{\bibfnamefont{H.}~\bibnamefont{Romberg}},
  \bibinfo{author}{\bibfnamefont{M.}~\bibnamefont{Alexander}},
  \bibnamefont{and} \bibinfo{author}{\bibfnamefont{M.}~\bibnamefont{Knupfer}},
  \bibinfo{journal}{J. Elect. Spect. and Rel. Phen.}
  \textbf{\bibinfo{volume}{66}}, \bibinfo{pages}{395} (\bibinfo{year}{1994}).

\bibitem[{\citenamefont{Zaanen et~al.}(1985)\citenamefont{Zaanen, Sawatzky, and
  Allen}}]{Zaanen85}
\bibinfo{author}{\bibfnamefont{J.}~\bibnamefont{Zaanen}},
  \bibinfo{author}{\bibfnamefont{G.~A.} \bibnamefont{Sawatzky}},
  \bibnamefont{and} \bibinfo{author}{\bibfnamefont{J.~W.} \bibnamefont{Allen}},
  \bibinfo{journal}{Phys. Rev. Lett.} \textbf{\bibinfo{volume}{55}},
  \bibinfo{pages}{418} (\bibinfo{year}{1985}).

\bibitem[{\citenamefont{Thomas et~al.}(1992)\citenamefont{Thomas, Rapkine,
  Cooper, Cheong, Cooper, Schneemeyer, and Waszczak}}]{Thomas92}
\bibinfo{author}{\bibfnamefont{G.~A.} \bibnamefont{Thomas}},
  \bibinfo{author}{\bibfnamefont{D.~H.} \bibnamefont{Rapkine}},
  \bibinfo{author}{\bibfnamefont{S.~L.} \bibnamefont{Cooper}},
  \bibinfo{author}{\bibfnamefont{S.-W.} \bibnamefont{Cheong}},
  \bibinfo{author}{\bibfnamefont{A.~S.} \bibnamefont{Cooper}},
  \bibinfo{author}{\bibfnamefont{L.~F.} \bibnamefont{Schneemeyer}},
  \bibnamefont{and} \bibinfo{author}{\bibfnamefont{J.~V.}
  \bibnamefont{Waszczak}}, \bibinfo{journal}{\prb}
  \textbf{\bibinfo{volume}{45}}, \bibinfo{pages}{2474} (\bibinfo{year}{1992}).

\bibitem[{\citenamefont{Basov and Timusk}(2005)}]{Basov05}
\bibinfo{author}{\bibfnamefont{D.~N.} \bibnamefont{Basov}} \bibnamefont{and}
  \bibinfo{author}{\bibfnamefont{T.}~\bibnamefont{Timusk}},
  \bibinfo{journal}{Rev. Mod. Phys.} \textbf{\bibinfo{volume}{77}},
  \bibinfo{pages}{721} (\bibinfo{year}{2005}).

\bibitem[{\citenamefont{Leggett}(1999)}]{Leggett99}
\bibinfo{author}{\bibfnamefont{A.~J.} \bibnamefont{Leggett}},
  \bibinfo{journal}{Proc. Natl. Acad. Sci. USA} \textbf{\bibinfo{volume}{96}},
  \bibinfo{pages}{8365} (\bibinfo{year}{1999}).

\bibitem[{\citenamefont{Lee et~al.}(2005)\citenamefont{Lee, Segawa, Li,
  Padilla, Dumm, Dordevic, Homes, Ando, and Basov}}]{Lee05}
\bibinfo{author}{\bibfnamefont{Y.~S.} \bibnamefont{Lee}},
  \bibinfo{author}{\bibfnamefont{K.}~\bibnamefont{Segawa}},
  \bibinfo{author}{\bibfnamefont{Z.~Q.} \bibnamefont{Li}},
  \bibinfo{author}{\bibfnamefont{W.~J.} \bibnamefont{Padilla}},
  \bibinfo{author}{\bibfnamefont{M.}~\bibnamefont{Dumm}},
  \bibinfo{author}{\bibfnamefont{S.~V.} \bibnamefont{Dordevic}},
  \bibinfo{author}{\bibfnamefont{C.~C.} \bibnamefont{Homes}},
  \bibinfo{author}{\bibfnamefont{Y.}~\bibnamefont{Ando}}, \bibnamefont{and}
  \bibinfo{author}{\bibfnamefont{D.~N.} \bibnamefont{Basov}},
  \bibinfo{journal}{\prb} \textbf{\bibinfo{volume}{72}},
  \bibinfo{pages}{054529} (\bibinfo{year}{2005}).

\bibitem[{\citenamefont{Quijada et~al.}(1995)\citenamefont{Quijada, Tanner,
  Chou, Johnston, and Cheong}}]{Quijada95}
\bibinfo{author}{\bibfnamefont{M.~A.} \bibnamefont{Quijada}},
  \bibinfo{author}{\bibfnamefont{D.~B.} \bibnamefont{Tanner}},
  \bibinfo{author}{\bibfnamefont{F.~C.} \bibnamefont{Chou}},
  \bibinfo{author}{\bibfnamefont{D.~C.} \bibnamefont{Johnston}},
  \bibnamefont{and} \bibinfo{author}{\bibfnamefont{S.~W.}
  \bibnamefont{Cheong}}, \bibinfo{journal}{\prb} \textbf{\bibinfo{volume}{52}},
  \bibinfo{pages}{15485} (\bibinfo{year}{1995}).

\bibitem[{\citenamefont{Ginder et~al.}(1988)\citenamefont{Ginder, Roe, Song, McCall,
  Gaines, Ehrenfreund, and Epstein}}]{Ginder88}
\bibinfo{author}{\bibfnamefont{J.~M.} \bibnamefont{Ginder}},
  \bibinfo{author}{\bibfnamefont{M.~G.} \bibnamefont{Roe}},
  \bibinfo{author}{\bibfnamefont{Y.} \bibnamefont{Song}},
  \bibinfo{author}{\bibfnamefont{R.~P.} \bibnamefont{McCall}},
  \bibinfo{author}{\bibfnamefont{J.~R.} \bibnamefont{Gaines}},
  \bibinfo{author}{\bibfnamefont{E.}~\bibnamefont{Ehrenfreund}},
  \bibnamefont{and} \bibinfo{author}{\bibfnamefont{A.~J.}
  \bibnamefont{Epstein}}, \bibinfo{journal}{\prb}
  \textbf{\bibinfo{volume}{37}}, \bibinfo{pages}{7506} (\bibinfo{year}{1988}).

\bibitem[{\citenamefont{Waku et~al.}(2004)\citenamefont{Waku, Katsufuji,
  Kohsaka, Sasagawa, Takagi, Kishida, Okamoto, Azuma, and Takano}}]{Waku04}
\bibinfo{author}{\bibfnamefont{K.}~\bibnamefont{Waku}},
  \bibinfo{author}{\bibfnamefont{T.}~\bibnamefont{Katsufuji}},
  \bibinfo{author}{\bibfnamefont{Y.}~\bibnamefont{Kohsaka}},
  \bibinfo{author}{\bibfnamefont{T.}~\bibnamefont{Sasagawa}},
  \bibinfo{author}{\bibfnamefont{H.}~\bibnamefont{Takagi}},
  \bibinfo{author}{\bibfnamefont{H.}~\bibnamefont{Kishida}},
  \bibinfo{author}{\bibfnamefont{H.}~\bibnamefont{Okamoto}},
  \bibinfo{author}{\bibfnamefont{M.}~\bibnamefont{Azuma}}, \bibnamefont{and}
  \bibinfo{author}{\bibfnamefont{M.}~\bibnamefont{Takano}},
  \bibinfo{journal}{\prb} \textbf{\bibinfo{volume}{70}},
  \bibinfo{pages}{134501} (\bibinfo{year}{2004}).

\bibitem[{\citenamefont{Abbamonte et~al.}(1999)\citenamefont{Abbamonte, Burns,
  Isaacs, Platzman, Miller, Cheong, and Klein}}]{Abbamonte99}
\bibinfo{author}{\bibfnamefont{P.}~\bibnamefont{Abbamonte}},
  \bibinfo{author}{\bibfnamefont{C.~A.} \bibnamefont{Burns}},
  \bibinfo{author}{\bibfnamefont{E.~D.} \bibnamefont{Isaacs}},
  \bibinfo{author}{\bibfnamefont{P.~M.} \bibnamefont{Platzman}},
  \bibinfo{author}{\bibfnamefont{L.~L.} \bibnamefont{Miller}},
  \bibinfo{author}{\bibfnamefont{S.~W.} \bibnamefont{Cheong}},
  \bibnamefont{and} \bibinfo{author}{\bibfnamefont{M.~V.} \bibnamefont{Klein}},
  \bibinfo{journal}{\prl} \textbf{\bibinfo{volume}{83}}, \bibinfo{pages}{860}
  (\bibinfo{year}{1999}).

\bibitem[{\citenamefont{Kim et~al.}(2002)\citenamefont{Kim, Hill, Burns,
  Wakimoto, Birgeneau, Casa, Gog, and Venkataraman}}]{Kim02}
\bibinfo{author}{\bibfnamefont{Y.~J.} \bibnamefont{Kim}},
  \bibinfo{author}{\bibfnamefont{J.~P.} \bibnamefont{Hill}},
  \bibinfo{author}{\bibfnamefont{C.~A.} \bibnamefont{Burns}},
  \bibinfo{author}{\bibfnamefont{S.}~\bibnamefont{Wakimoto}},
  \bibinfo{author}{\bibfnamefont{R.~J.} \bibnamefont{Birgeneau}},
  \bibinfo{author}{\bibfnamefont{D.}~\bibnamefont{Casa}},
  \bibinfo{author}{\bibfnamefont{T.}~\bibnamefont{Gog}}, \bibnamefont{and}
  \bibinfo{author}{\bibfnamefont{C.~T.} \bibnamefont{Venkataraman}},
  \bibinfo{journal}{\prl} \textbf{\bibinfo{volume}{89}},
  \bibinfo{pages}{177003} (\bibinfo{year}{2002}).

\bibitem[{\citenamefont{Collart et~al.}(2006)\citenamefont{Collart, Shukla,
  Rueff, Leininger, Ishii, Jarrige, Cai, Cheong, and Dhalenne}}]{Collart06}
\bibinfo{author}{\bibfnamefont{E.}~\bibnamefont{Collart}},
  \bibinfo{author}{\bibfnamefont{A.}~\bibnamefont{Shukla}},
  \bibinfo{author}{\bibfnamefont{J.-P.} \bibnamefont{Rueff}},
  \bibinfo{author}{\bibfnamefont{P.}~\bibnamefont{Leininger}},
  \bibinfo{author}{\bibfnamefont{H.}~\bibnamefont{Ishii}},
  \bibinfo{author}{\bibfnamefont{I.}~\bibnamefont{Jarrige}},
  \bibinfo{author}{\bibfnamefont{Y.~Q.} \bibnamefont{Cai}},
  \bibinfo{author}{\bibfnamefont{S.-W.} \bibnamefont{Cheong}},
  \bibnamefont{and} \bibinfo{author}{\bibfnamefont{G.}~\bibnamefont{Dhalenne}},
  \bibinfo{journal}{\prl} \textbf{\bibinfo{volume}{96}},
  \bibinfo{pages}{157004} (\bibinfo{year}{2006}).

\bibitem[{\citenamefont{Lu et~al.}(2006)\citenamefont{Lu, Hancock,
  Chabot-Couture, Ishii, Vajk, Yu, Mizuki, Casa, Gog, and Greven}}]{Lu06}
\bibinfo{author}{\bibfnamefont{L.}~\bibnamefont{Lu}},
  \bibinfo{author}{\bibfnamefont{J.~N.} \bibnamefont{Hancock}},
  \bibinfo{author}{\bibfnamefont{G.}~\bibnamefont{Chabot-Couture}},
  \bibinfo{author}{\bibfnamefont{K.}~\bibnamefont{Ishii}},
  \bibinfo{author}{\bibfnamefont{O.~P.} \bibnamefont{Vajk}},
  \bibinfo{author}{\bibfnamefont{G.}~\bibnamefont{Yu}},
  \bibinfo{author}{\bibfnamefont{J.}~\bibnamefont{Mizuki}},
  \bibinfo{author}{\bibfnamefont{D.}~\bibnamefont{Casa}},
  \bibinfo{author}{\bibfnamefont{T.}~\bibnamefont{Gog}}, \bibnamefont{and}
  \bibinfo{author}{\bibfnamefont{M.}~\bibnamefont{Greven}},
  \bibinfo{journal}{\prb} \textbf{\bibinfo{volume}{74}},
  \bibinfo{pages}{224509} (\bibinfo{year}{2006}).

\bibitem[{\citenamefont{Ellis et~al.}(2008)\citenamefont{Ellis, Hill, Wakimoto,
  Birgeneau, Casa, Gog, and Kim}}]{Ellis08}
\bibinfo{author}{\bibfnamefont{D.~S.}~\bibnamefont{Ellis}},
  \bibinfo{author}{\bibfnamefont{J.~P.} \bibnamefont{Hill}},
  \bibinfo{author}{\bibfnamefont{S.}~\bibnamefont{Wakimoto}},
  \bibinfo{author}{\bibfnamefont{R.~J.} \bibnamefont{Birgeneau}},
  \bibinfo{author}{\bibfnamefont{D.}~\bibnamefont{Casa}},
  \bibinfo{author}{\bibfnamefont{T.}~\bibnamefont{Gog}}, \bibnamefont{and}
  \bibinfo{author}{\bibfnamefont{Y.-J.} \bibnamefont{Kim}},
  \bibinfo{journal}{\prb} \textbf{\bibinfo{volume}{77}},
  \bibinfo{pages}{060501} (\bibinfo{year}{2008}).

\bibitem[{\citenamefont{Kim et~al.}(2004)\citenamefont{Kim, Hill, Komiya, Ando,
  Casa, Gog, and Venkataraman}}]{Kim04}
\bibinfo{author}{\bibfnamefont{Y.~J.} \bibnamefont{Kim}},
  \bibinfo{author}{\bibfnamefont{J.~P.} \bibnamefont{Hill}},
  \bibinfo{author}{\bibfnamefont{S.}~\bibnamefont{Komiya}},
  \bibinfo{author}{\bibfnamefont{Y.}~\bibnamefont{Ando}},
  \bibinfo{author}{\bibfnamefont{D.}~\bibnamefont{Casa}},
  \bibinfo{author}{\bibfnamefont{T.}~\bibnamefont{Gog}}, \bibnamefont{and}
  \bibinfo{author}{\bibfnamefont{C.~T.} \bibnamefont{Venkataraman}},
  \bibinfo{journal}{\prb} \textbf{\bibinfo{volume}{70}},
  \bibinfo{pages}{094524} (\bibinfo{year}{2004}).

\bibitem[{\citenamefont{Wakimoto et~al.}(2005)\citenamefont{Wakimoto, Kim, Kim,
  Zhang, Gog, and Birgeneau}}]{Wakimoto05}
\bibinfo{author}{\bibfnamefont{S.}~\bibnamefont{Wakimoto}},
  \bibinfo{author}{\bibfnamefont{Y.~J.} \bibnamefont{Kim}},
  \bibinfo{author}{\bibfnamefont{H.}~\bibnamefont{Kim}},
  \bibinfo{author}{\bibfnamefont{H.}~\bibnamefont{Zhang}},
  \bibinfo{author}{\bibfnamefont{T.}~\bibnamefont{Gog}}, \bibnamefont{and}
  \bibinfo{author}{\bibfnamefont{R.~J.} \bibnamefont{Birgeneau}},
  \bibinfo{journal}{\prb} \textbf{\bibinfo{volume}{72}},
  \bibinfo{pages}{224508} (\bibinfo{year}{2005}).

\bibitem[{\citenamefont{Kim et~al.}(2009)\citenamefont{Kim, Ellis, Zhang, Kim,
  Hill, Chou, Gog, and Casa}}]{Kim09}
\bibinfo{author}{\bibfnamefont{J.}~\bibnamefont{Kim}},
  \bibinfo{author}{\bibfnamefont{D.~S.} \bibnamefont{Ellis}},
  \bibinfo{author}{\bibfnamefont{H.}~\bibnamefont{Zhang}},
  \bibinfo{author}{\bibfnamefont{Y.-J.} \bibnamefont{Kim}},
  \bibinfo{author}{\bibfnamefont{J.~P.} \bibnamefont{Hill}},
  \bibinfo{author}{\bibfnamefont{F.~C.} \bibnamefont{Chou}},
  \bibinfo{author}{\bibfnamefont{T.}~\bibnamefont{Gog}}, \bibnamefont{and}
  \bibinfo{author}{\bibfnamefont{D.}~\bibnamefont{Casa}},
  \bibinfo{journal}{Phys. Rev. B.} \textbf{\bibinfo{volume}{79}},
  \bibinfo{pages}{094525} (\bibinfo{year}{2009}).

\bibitem[{\citenamefont{Ament et~al.}(2007)\citenamefont{Ament, Forte, and
  van~den Brink}}]{Ament07}
\bibinfo{author}{\bibfnamefont{L.~J.~P.} \bibnamefont{Ament}},
  \bibinfo{author}{\bibfnamefont{F.}~\bibnamefont{Forte}}, \bibnamefont{and}
  \bibinfo{author}{\bibfnamefont{J.}~\bibnamefont{van~den Brink}},
  \bibinfo{journal}{Phys. Rev. B.} \textbf{\bibinfo{volume}{75}},
  \bibinfo{pages}{115118} (\bibinfo{year}{2007}).

\bibitem[{\citenamefont{Tsutsui et~al.}(1999)\citenamefont{Tsutsui, Tohyama,
  and Maekawa}}]{Tsutsui99}
\bibinfo{author}{\bibfnamefont{K.}~\bibnamefont{Tsutsui}},
  \bibinfo{author}{\bibfnamefont{T.}~\bibnamefont{Tohyama}}, \bibnamefont{and}
  \bibinfo{author}{\bibfnamefont{S.}~\bibnamefont{Maekawa}},
  \bibinfo{journal}{\prl} \textbf{\bibinfo{volume}{83}}, \bibinfo{pages}{3705}
  (\bibinfo{year}{1999}).

\bibitem[{\citenamefont{Markiewicz and Bansil}(2006)}]{Markiewicz06b}
\bibinfo{author}{\bibfnamefont{R.~S.} \bibnamefont{Markiewicz}}
  \bibnamefont{and} \bibinfo{author}{\bibfnamefont{A.}~\bibnamefont{Bansil}},
  \bibinfo{journal}{Phys. Rev. Lett.} \textbf{\bibinfo{volume}{96}},
  \bibinfo{pages}{107005} (\bibinfo{year}{2006}).

\bibitem[{\citenamefont{Ghiringhelli et~al.}(2004)\citenamefont{Ghiringhelli,
  Brookes, Annese, Berger, Dallera, Grioni, Perfetti, Tagliaferri, and
  Braicovich}}]{Ghiringhelli04}
\bibinfo{author}{\bibfnamefont{G.}~\bibnamefont{Ghiringhelli}},
  \bibinfo{author}{\bibfnamefont{N.~B.} \bibnamefont{Brookes}},
  \bibinfo{author}{\bibfnamefont{E.}~\bibnamefont{Annese}},
  \bibinfo{author}{\bibfnamefont{H.}~\bibnamefont{Berger}},
  \bibinfo{author}{\bibfnamefont{C.}~\bibnamefont{Dallera}},
  \bibinfo{author}{\bibfnamefont{M.}~\bibnamefont{Grioni}},
  \bibinfo{author}{\bibfnamefont{L.}~\bibnamefont{Perfetti}},
  \bibinfo{author}{\bibfnamefont{A.}~\bibnamefont{Tagliaferri}},
  \bibnamefont{and}
  \bibinfo{author}{\bibfnamefont{L.}~\bibnamefont{Braicovich}},
  \bibinfo{journal}{\prl} \textbf{\bibinfo{volume}{92}},
  \bibinfo{pages}{117406} (\bibinfo{year}{2004}).

\bibitem[{\citenamefont{Wakimoto et~al.}(2009)\citenamefont{Wakimoto, Kimura,
  Ishii, Ikeuchi, Adachi, Fujita, Kakurai, Koike, Mizuki, Noda
  et~al.}}]{Wakimoto09}
\bibinfo{author}{\bibfnamefont{S.}~\bibnamefont{Wakimoto}},
  \bibinfo{author}{\bibfnamefont{H.}~\bibnamefont{Kimura}},
  \bibinfo{author}{\bibfnamefont{K.}~\bibnamefont{Ishii}},
  \bibinfo{author}{\bibfnamefont{K.}~\bibnamefont{Ikeuchi}},
  \bibinfo{author}{\bibfnamefont{T.}~\bibnamefont{Adachi}},
  \bibinfo{author}{\bibfnamefont{M.}~\bibnamefont{Fujita}},
  \bibinfo{author}{\bibfnamefont{K.}~\bibnamefont{Kakurai}},
  \bibinfo{author}{\bibfnamefont{Y.}~\bibnamefont{Koike}},
  \bibinfo{author}{\bibfnamefont{J.}~\bibnamefont{Mizuki}},
  \bibinfo{author}{\bibfnamefont{Y.}~\bibnamefont{Noda}}, \bibnamefont{et~al.},
  \bibinfo{journal}{\prl} \textbf{\bibinfo{volume}{102}},
  \bibinfo{pages}{157001} (\bibinfo{year}{2009}).

\bibitem[{\citenamefont{Kim et~al.}(2008)\citenamefont{Kim, Kagedan, Gu,
  Nelson, and Kim}}]{Kim08}
\bibinfo{author}{\bibfnamefont{J.}~\bibnamefont{Kim}},
  \bibinfo{author}{\bibfnamefont{A.}~\bibnamefont{Kagedan}},
  \bibinfo{author}{\bibfnamefont{G.~D.} \bibnamefont{Gu}},
  \bibinfo{author}{\bibfnamefont{C.~S.} \bibnamefont{Nelson}},
  \bibnamefont{and} \bibinfo{author}{\bibfnamefont{Y.-J.} \bibnamefont{Kim}},
  \bibinfo{journal}{\prb} \textbf{\bibinfo{volume}{77}},
  \bibinfo{pages}{180513} (\bibinfo{year}{2008}).

\bibitem[{\citenamefont{Huotari et~al.}(2005)\citenamefont{Huotari, Vanko,
  Albergamo, Ponchut, Graafsma, Henriquet, Verbeni, and Monaco}}]{Huotari05}
\bibinfo{author}{\bibfnamefont{S.}~\bibnamefont{Huotari}},
  \bibinfo{author}{\bibfnamefont{G.}~\bibnamefont{Vanko}},
  \bibinfo{author}{\bibfnamefont{F.}~\bibnamefont{Albergamo}},
  \bibinfo{author}{\bibfnamefont{C.}~\bibnamefont{Ponchut}},
  \bibinfo{author}{\bibfnamefont{H.}~\bibnamefont{Graafsma}},
  \bibinfo{author}{\bibfnamefont{C.}~\bibnamefont{Henriquet}},
  \bibinfo{author}{\bibfnamefont{R.}~\bibnamefont{Verbeni}}, \bibnamefont{and}
  \bibinfo{author}{\bibfnamefont{G.}~\bibnamefont{Monaco}},
  \bibinfo{journal}{J. of Synchr. Rad.} \textbf{\bibinfo{volume}{12}},
  \bibinfo{pages}{467} (\bibinfo{year}{2005}).

\bibitem[{\citenamefont{Eckstein et~al.}(2007)\citenamefont{Eckstein, Kollar,
  and Vollhardt}}]{Eckstein07}
\bibinfo{author}{\bibfnamefont{M.}~\bibnamefont{Eckstein}},
  \bibinfo{author}{\bibfnamefont{M.}~\bibnamefont{Kollar}}, \bibnamefont{and}
  \bibinfo{author}{\bibfnamefont{D.}~\bibnamefont{Vollhardt}},
  \bibinfo{journal}{Journal of Low Temperature Physics}
  \textbf{\bibinfo{volume}{147}}, \bibinfo{pages}{279} (\bibinfo{year}{2007}).

\bibitem[{\citenamefont{Ino et~al.}(1997)\citenamefont{Ino, Mizokawa, Fujimori,
  Tamasaku, Eisaki, Uchida, Kimura, Sasagawa, and Kishio}}]{Ino97}
\bibinfo{author}{\bibfnamefont{A.}~\bibnamefont{Ino}},
  \bibinfo{author}{\bibfnamefont{T.}~\bibnamefont{Mizokawa}},
  \bibinfo{author}{\bibfnamefont{A.}~\bibnamefont{Fujimori}},
  \bibinfo{author}{\bibfnamefont{K.}~\bibnamefont{Tamasaku}},
  \bibinfo{author}{\bibfnamefont{H.}~\bibnamefont{Eisaki}},
  \bibinfo{author}{\bibfnamefont{S.}~\bibnamefont{Uchida}},
  \bibinfo{author}{\bibfnamefont{T.}~\bibnamefont{Kimura}},
  \bibinfo{author}{\bibfnamefont{T.}~\bibnamefont{Sasagawa}}, \bibnamefont{and}
  \bibinfo{author}{\bibfnamefont{K.}~\bibnamefont{Kishio}},
  \bibinfo{journal}{\prl} \textbf{\bibinfo{volume}{79}}, \bibinfo{pages}{2101}
  (\bibinfo{year}{1997}).

\bibitem[{\citenamefont{Ellis et~al.}(2010)\citenamefont{Ellis, Kim, Hill,
  Wakimoto, Birgeneau, Shvyd'ko, Casa, Gog, Ishii, Ikeuchi et~al.}}]{Ellis10}
\bibinfo{author}{\bibfnamefont{D.~S.} \bibnamefont{Ellis}},
  \bibinfo{author}{\bibfnamefont{J.}~\bibnamefont{Kim}},
  \bibinfo{author}{\bibfnamefont{J.~P.} \bibnamefont{Hill}},
  \bibinfo{author}{\bibfnamefont{S.}~\bibnamefont{Wakimoto}},
  \bibinfo{author}{\bibfnamefont{R.~J.} \bibnamefont{Birgeneau}},
  \bibinfo{author}{\bibfnamefont{Y.}~\bibnamefont{Shvyd'ko}},
  \bibinfo{author}{\bibfnamefont{D.}~\bibnamefont{Casa}},
  \bibinfo{author}{\bibfnamefont{T.}~\bibnamefont{Gog}},
  \bibinfo{author}{\bibfnamefont{K.}~\bibnamefont{Ishii}},
  \bibinfo{author}{\bibfnamefont{K.}~\bibnamefont{Ikeuchi}},
  \bibnamefont{et~al.}, \bibinfo{journal}{\prb} \textbf{\bibinfo{volume}{81}},
  \bibinfo{pages}{085124} (\bibinfo{year}{2010}).

\bibitem[{\citenamefont{Nomura and Igarashi}(2005)}]{Nomura05}
\bibinfo{author}{\bibfnamefont{T.}~\bibnamefont{Nomura}} \bibnamefont{and}
  \bibinfo{author}{\bibfnamefont{J.~I.}~\bibnamefont{Igarashi}},
  \bibinfo{journal}{\prb} \textbf{\bibinfo{volume}{71}},
  \bibinfo{pages}{035110} (\bibinfo{year}{2005}).

\bibitem[{\citenamefont{Kim et~al.}(2010)\citenamefont{Kim, Hill, Yamaguchi,
  Gog, and Casa}}]{Kim10}
\bibinfo{author}{\bibfnamefont{Y.-J.} \bibnamefont{Kim}},
  \bibinfo{author}{\bibfnamefont{J.~P.}~\bibnamefont{Hill}},
  \bibinfo{author}{\bibfnamefont{H.}~\bibnamefont{Yamaguchi}},
  \bibinfo{author}{\bibfnamefont{T.}~\bibnamefont{Gog}}, \bibnamefont{and}
  \bibinfo{author}{\bibfnamefont{D.}~\bibnamefont{Casa}},
  \bibinfo{journal}{\prb} \textbf{\bibinfo{volume}{81}},
  \bibinfo{pages}{195202} (\bibinfo{year}{2010}).

\bibitem[{\citenamefont{Tobin et~al.}(1992)\citenamefont{Tobin, Olson, Gu, Liu,
  Solal, Fluss, Howell, O'Brien, Radousky, and Sterne}}]{Tobin92}
\bibinfo{author}{\bibfnamefont{J.~G.} \bibnamefont{Tobin}},
  \bibinfo{author}{\bibfnamefont{C.~G.} \bibnamefont{Olson}},
  \bibinfo{author}{\bibfnamefont{C.}~\bibnamefont{Gu}},
  \bibinfo{author}{\bibfnamefont{J.~Z.} \bibnamefont{Liu}},
  \bibinfo{author}{\bibfnamefont{F.~R.} \bibnamefont{Solal}},
  \bibinfo{author}{\bibfnamefont{M.~J.} \bibnamefont{Fluss}},
  \bibinfo{author}{\bibfnamefont{R.~H.} \bibnamefont{Howell}},
  \bibinfo{author}{\bibfnamefont{J.~C.} \bibnamefont{O'Brien}},
  \bibinfo{author}{\bibfnamefont{H.~B.} \bibnamefont{Radousky}},
  \bibnamefont{and} \bibinfo{author}{\bibfnamefont{P.~A.}
  \bibnamefont{Sterne}}, \bibinfo{journal}{\prb} \textbf{\bibinfo{volume}{45}},
  \bibinfo{pages}{5563} (\bibinfo{year}{1992}).

\bibitem[{\citenamefont{Olson et~al.}(1995)\citenamefont{Olson, Tobin, Waddill,
  Lynch, and Liu}}]{Olson95}
\bibinfo{author}{\bibfnamefont{C.~G.} \bibnamefont{Olson}},
  \bibinfo{author}{\bibfnamefont{J.~G.} \bibnamefont{Tobin}},
  \bibinfo{author}{\bibfnamefont{G.~D.} \bibnamefont{Waddill}},
  \bibinfo{author}{\bibfnamefont{D.~W.} \bibnamefont{Lynch}}, \bibnamefont{and}
  \bibinfo{author}{\bibfnamefont{J.~Z.} \bibnamefont{Liu}},
  \bibinfo{journal}{J. Phys. Chem. Solids} \textbf{\bibinfo{volume}{56}},
  \bibinfo{pages}{1879} (\bibinfo{year}{1995}).

\bibitem[{\citenamefont{Pothuizen et~al.}(1997)\citenamefont{Pothuizen, Eder,
  Hien, Matoba, Menovsky, and Sawatzky}}]{Pothuizen97}
\bibinfo{author}{\bibfnamefont{J.~J.~M.} \bibnamefont{Pothuizen}},
  \bibinfo{author}{\bibfnamefont{R.}~\bibnamefont{Eder}},
  \bibinfo{author}{\bibfnamefont{N.~T.} \bibnamefont{Hien}},
  \bibinfo{author}{\bibfnamefont{M.}~\bibnamefont{Matoba}},
  \bibinfo{author}{\bibfnamefont{A.~A.} \bibnamefont{Menovsky}},
  \bibnamefont{and} \bibinfo{author}{\bibfnamefont{G.~A.}
  \bibnamefont{Sawatzky}}, \bibinfo{journal}{\prl}
  \textbf{\bibinfo{volume}{78}}, \bibinfo{pages}{717} (\bibinfo{year}{1997}).

\bibitem[{\citenamefont{Durr et~al.}(2000)\citenamefont{Durr, Legner, Hayn,
  Borisenko, Hu, Theresiak, Knupfer, Golden, Fink, Ronning et~al.}}]{Durr00}
\bibinfo{author}{\bibfnamefont{C.}~\bibnamefont{Durr}},
  \bibinfo{author}{\bibfnamefont{S.}~\bibnamefont{Legner}},
  \bibinfo{author}{\bibfnamefont{R.}~\bibnamefont{Hayn}},
  \bibinfo{author}{\bibfnamefont{S.~V.} \bibnamefont{Borisenko}},
  \bibinfo{author}{\bibfnamefont{Z.}~\bibnamefont{Hu}},
  \bibinfo{author}{\bibfnamefont{A.}~\bibnamefont{Theresiak}},
  \bibinfo{author}{\bibfnamefont{M.}~\bibnamefont{Knupfer}},
  \bibinfo{author}{\bibfnamefont{M.~S.} \bibnamefont{Golden}},
  \bibinfo{author}{\bibfnamefont{J.}~\bibnamefont{Fink}},
  \bibinfo{author}{\bibfnamefont{F.}~\bibnamefont{Ronning}},
  \bibnamefont{et~al.}, \bibinfo{journal}{\prb} \textbf{\bibinfo{volume}{63}},
  \bibinfo{pages}{014505} (\bibinfo{year}{2000}).

\bibitem[{\citenamefont{Wagner et~al.}(1991)\citenamefont{Wagner, Hanke, and
  Scalapino}}]{Wagner91}
\bibinfo{author}{\bibfnamefont{J.}~\bibnamefont{Wagner}},
  \bibinfo{author}{\bibfnamefont{W.}~\bibnamefont{Hanke}}, \bibnamefont{and}
  \bibinfo{author}{\bibfnamefont{D.~J.} \bibnamefont{Scalapino}},
  \bibinfo{journal}{\prb} \textbf{\bibinfo{volume}{43}}, \bibinfo{pages}{10517}
  (\bibinfo{year}{1991}).

\bibitem[{\citenamefont{Fleck et~al.}(2001)\citenamefont{Fleck, Lichtenstein,
  and Oles}}]{Fleck01}
\bibinfo{author}{\bibfnamefont{M.}~\bibnamefont{Fleck}},
  \bibinfo{author}{\bibfnamefont{A.~I.} \bibnamefont{Lichtenstein}},
  \bibnamefont{and} \bibinfo{author}{\bibfnamefont{A.~M.} \bibnamefont{Oles}},
  \bibinfo{journal}{\prb} \textbf{\bibinfo{volume}{64}},
  \bibinfo{pages}{134528} (\bibinfo{year}{2001}).

\bibitem[{\citenamefont{Peets et~al.}(2009)\citenamefont{Peets, Hawthorn, Shen,
  Kim, Ellis, Zhang, Komiya, Ando, Sawatzky, Liang et~al.}}]{Peets09}
\bibinfo{author}{\bibfnamefont{D.~C.} \bibnamefont{Peets}},
  \bibinfo{author}{\bibfnamefont{D.~G.} \bibnamefont{Hawthorn}},
  \bibinfo{author}{\bibfnamefont{K.~M.} \bibnamefont{Shen}},
  \bibinfo{author}{\bibfnamefont{Y.-J.} \bibnamefont{Kim}},
  \bibinfo{author}{\bibfnamefont{D.~S.} \bibnamefont{Ellis}},
  \bibinfo{author}{\bibfnamefont{H.}~\bibnamefont{Zhang}},
  \bibinfo{author}{\bibfnamefont{S.}~\bibnamefont{Komiya}},
  \bibinfo{author}{\bibfnamefont{Y.}~\bibnamefont{Ando}},
  \bibinfo{author}{\bibfnamefont{G.~A.} \bibnamefont{Sawatzky}},
  \bibinfo{author}{\bibfnamefont{R.}~\bibnamefont{Liang}},
  \bibnamefont{et~al.}, \bibinfo{journal}{Phys. Rev. Lett.}
  \textbf{\bibinfo{volume}{103}}, \bibinfo{pages}{087402}
  (\bibinfo{year}{2009}).

\bibitem[{\citenamefont{de'Medici et~al.}(2009)\citenamefont{de'Medici, Wang,
  Capone, and Millis}}]{Medici09}
\bibinfo{author}{\bibfnamefont{L.}~\bibnamefont{de'Medici}},
  \bibinfo{author}{\bibfnamefont{X.}~\bibnamefont{Wang}},
  \bibinfo{author}{\bibfnamefont{M.}~\bibnamefont{Capone}}, \bibnamefont{and}
  \bibinfo{author}{\bibfnamefont{A.~J.} \bibnamefont{Millis}},
  \bibinfo{journal}{\prb} \textbf{\bibinfo{volume}{80}},
  \bibinfo{pages}{054501} (\bibinfo{year}{2009}).

\bibitem[{\citenamefont{Zhang and Ng}(1998)}]{Zhang98}
\bibinfo{author}{\bibfnamefont{F.~C.} \bibnamefont{Zhang}} \bibnamefont{and}
  \bibinfo{author}{\bibfnamefont{K.~K.} \bibnamefont{Ng}},
  \bibinfo{journal}{\prb} \textbf{\bibinfo{volume}{58}}, \bibinfo{pages}{13520}
  (\bibinfo{year}{1998}).

\bibitem[{\citenamefont{Takahashi et~al.}(2009)\citenamefont{Takahashi,
  Igarashi, and Semba}}]{Takahashi09}
\bibinfo{author}{\bibfnamefont{M.}~\bibnamefont{Takahashi}},
  \bibinfo{author}{\bibfnamefont{J.}~\bibnamefont{Igarashi}}, \bibnamefont{and}
  \bibinfo{author}{\bibfnamefont{T.}~\bibnamefont{Semba}}, \bibinfo{journal}{J.
  Phys.: Condens. Matter} \textbf{\bibinfo{volume}{21}},
  \bibinfo{pages}{064236} (\bibinfo{year}{2009}).

\bibitem[{\citenamefont{Schuster et~al.}(2009)\citenamefont{Schuster, Pyon,
  Knupfer, Fink, Azuma, Takano, Takagi, and Buchner}}]{Schuster09}
\bibinfo{author}{\bibfnamefont{R.}~\bibnamefont{Schuster}},
  \bibinfo{author}{\bibfnamefont{S.}~\bibnamefont{Pyon}},
  \bibinfo{author}{\bibfnamefont{M.}~\bibnamefont{Knupfer}},
  \bibinfo{author}{\bibfnamefont{J.}~\bibnamefont{Fink}},
  \bibinfo{author}{\bibfnamefont{M.}~\bibnamefont{Azuma}},
  \bibinfo{author}{\bibfnamefont{M.}~\bibnamefont{Takano}},
  \bibinfo{author}{\bibfnamefont{H.}~\bibnamefont{Takagi}}, \bibnamefont{and}
  \bibinfo{author}{\bibfnamefont{B.}~\bibnamefont{Buchner}},
  \bibinfo{journal}{\prb} \textbf{\bibinfo{volume}{79}},
  \bibinfo{pages}{214517} (\bibinfo{year}{2009}).

\bibitem[{\citenamefont{Hiraka et~al.}(2009)\citenamefont{Hiraka, Matsumura,
  Nishihata, Mizuki, and Yamada}}]{Hiraka09}
\bibinfo{author}{\bibfnamefont{H.}~\bibnamefont{Hiraka}},
  \bibinfo{author}{\bibfnamefont{D.}~\bibnamefont{Matsumura}},
  \bibinfo{author}{\bibfnamefont{Y.}~\bibnamefont{Nishihata}},
  \bibinfo{author}{\bibfnamefont{J.}~\bibnamefont{Mizuki}}, \bibnamefont{and}
  \bibinfo{author}{\bibfnamefont{K.}~\bibnamefont{Yamada}},
  \bibinfo{journal}{\prl} \textbf{\bibinfo{volume}{102}},
  \bibinfo{pages}{037002} (\bibinfo{year}{2009}).


\bibitem{Lyons88}
\bibinfo{author}{\bibfnamefont{K.~B.}~\bibnamefont{Lyons}},
\bibinfo{author}{\bibfnamefont{P.~A.}~\bibnamefont{Fleury}},
\bibinfo{author}{\bibfnamefont{J.~P.}~\bibnamefont{Remeika}},
\bibinfo{author}{\bibfnamefont{A.~S.}~\bibnamefont{Cooper}},
\bibinfo{author}{\bibfnamefont{T.~J.}~\bibnamefont{Negran}},
\bibinfo{journal}{\prb} \textbf{\bibinfo{volume}{37}},
\bibinfo{pages}{2353} (\bibinfo{year}{1988}).


\end{thebibliography}
\end{document}